\documentclass[a4paper,12pt]{article}
\usepackage{amssymb}
\usepackage{amsmath}
\usepackage{amsfonts}
\usepackage{amsthm}
\usepackage{graphicx}
\usepackage{epstopdf}
\usepackage{mathtools}
\usepackage{enumerate}
\usepackage{graphicx}
\usepackage{fullpage}
\usepackage{hyperref}
\usepackage{verbatim}
\DeclarePairedDelimiter{\ceil}{\lceil}{\rceil}
\newtheorem*{theorem*}{Theorem}

\newtheorem{thm}{Theorem}[section]
\newtheorem{prop}[thm]{Proposition}
\newtheorem{lemma}[thm]{Lemma}
\newtheorem{defn}[thm]{Definition}

\newtheorem{coro}[thm]{Corollary}

\begin{document}
	
	\begin{center}
		\textsc{\large\bf{{A SURVEY ON PRODUCT CODES AND 2-D CODES}}}
	\end{center}

	\begin{center}
		\textbf{Amajit Sarma}
	\end{center}

	\vspace{0.1cm}
	
	\begin{abstract}
	    One of the simplest way of combining codes to form new codes is to take their direct product. Direct product of cyclic codes and various generalizations have been studied for many years. In this note, we survey cyclic product codes, direct product of various generalizations of cyclic codes and their properties.
	\end{abstract}

	\section{Introduction}
		
		\paragraph*{} One of the most common method of combining existing linear codes to form new ones is to take the direct product of the codes. Given an $[n_1, k_1, d_1]$-code $C_1$ and an $[n_2, k_2, d_2]$-code $C_2$ over the finite field $\mathbb{F}_{q}$, where $q$ is a power of a prime $p$. Their direct product $C_{1} \otimes C_{2}$ is an $[n_{1}n_{2}, k_{1}k_{2}, d_{1}d_{2}]$-code over $\mathbb{F}_{q}$ whose codeword consists of all $n_{1} \times n_{2}$ arrays in which the columns are codewords of $C_{1}$ and rows are codewords of $C_{2}$. The direct product of codes, commonly referred to as product codes, have poor minimum distance. But there are some applications that use rectangles as codewords instead of vectors and these product codes and their generalizations are very useful here. The aim of this article is to survey various generalizations of product codes and study their properties. Our primary focus in the article is to compile together results on the cyclic product codes, two dimensional cyclic codes, quasi-cyclic product codes, 2-D skew-cyclic codes and 2-D skew-constacyclic codes.
	
		\paragraph*{} If $C_{1}$ and $C_{2}$ are cyclic codes of length $n_1$ and $n_2$ respectively, then the product code $C_{1} \otimes C_{2}$ is invariant under the cyclic permutation of all rows simultaneously or all columns simultaneously of its codewords. A codeword in $C_{1} \otimes C_{2}$
		\[ \begin{bmatrix}
		a_{0,0} & a_{0,1} & \dots & a_{0, n_{2}-1}\\
		a_{1,0} & a_{1,1} & \dots & a_{1, n_{2}-1}\\
		\vdots &\vdots & \vdots & \vdots\\
		a_{n_{1}-1, 0} & a_{n_{1}-1,1} & \dots & a_{n_{1}-1, n_{2}-1}
		\end{bmatrix}
		\]
		can be represented as a polynomial $f(x,y)=\sum_{i=0}^{n_{1}-1} \sum_{j=0}^{n_{2}-1} a_{i,j} x^i y^j$ in two variables. If we assume $x^{n_{1}}=1$ and $y^{n_{2}}=1$, then in terms of polynomials we have $xf(x,y), yf(x,y) \in C_{1} \otimes C_{2}$. In case $n_{1}$ and $n_{2}$ are relatively prime, using Chinese remainder theorem, a cyclic structure to $C_{1} \otimes C_{2}$ can be given as stated in the following theorem proved in \cite{peterson1961error}.
		\begin{thm}[\cite{peterson1961error}, Ch. $1$, Theorem $1$]
			If $C_{1}$ and $C_{2}$ are cyclic codes and $\gcd(n_{1},n_{2})=1$, then $C_{1} \otimes C_{2}$ is also cyclic.
		\end{thm}
		Section $2$ focuses on cyclic product codes. After giving a cyclic structure to $C_{1} \otimes C_{2}$ in Theorem $1.1$, Theorem $2.2$ gives the generator polynomial of the product code as presented in \cite{burton1965cyclic}. Section $2.1$ gives an application of cyclic product codes. Determining minimum distance of linear codes is a hard problem. Theorem $2.5$, $2.7$ and $2.8$ state the various bounds on the minimum distance of constituent cyclic codes using cyclic product codes as determined in \cite{zeh2013generalizing}.
		
		\paragraph*{} Another important parameter of linear codes is their generalized Hamming weights, which determine their usefulness in cryptography. Section $3$ states results on the generalized Hamming weights of product codes, proved in \cite{wei1993on}. If the constituent codes satisfy chain conditions then bounds on the generalized Hamming weights of product codes can be given as stated in Theorem $3.4$.
		
		\paragraph*{} Two-dimensional cyclic codes are another variant of cyclic codes with codewords as matrices. The paper \cite{guneri2012a relation} shows that these codes have relation to quasi-cyclic codes as shown in Section $4.1$. Generator matrix is a concise way of representing linear codes and thus is important to know. Section $4.2$ deals with generator matrix of two-dimensional cyclic codes as in \cite{sepasdar2017generator}. Some two-dimensional cyclic codes corresponding to particular ideals as discussed in \cite{sepasdar2016characterizations} is surveyed in Section $4.3$. Section $4.4$ is slightly different from rest of the section as it requires knowledge of the algebraic curves over finite fields especially the Artin-Schreier curves. Using the information on number of solutions of Arin-Schreier curve over $\mathbb{F}_{q^m}$ in $\mathbb{F}_{q^m} \times \mathbb{F}_{q^m}$, 2-D cyclic codes are constructed and bound on the minimum distance is determined in \cite{guneri2004artin}.
		
		\paragraph*{} Section $5$ generalizes the ideas of \cite{burton1965cyclic} to survey another variant of cyclic product codes called quasi-cyclic product codes. Quasi-cyclic codes form an important family of linear codes. Many good codes belong to this family of codes. Thus, it is interesting to study properties of quasi-cyclic product codes, as done in \cite{zeh2016spectral}. The results on the generator polynomials are stated. Section $6$ focuses on 2-D skew cyclic codes over various rings and state the known results on these codes. Section $7$ deals with 2-D Constacyclic codes studied in \cite{mostafanasab20152-dskew}. Using CSS construction, researchers have always obtained good families of quantum codes from linear codes. The paper \cite{la guardia2012asymmetric} extends this idea to obtain quantum product codes using linear product codes from CSS construction. Theorem $8.3$ and subsequent results taken from \cite{la guardia2012asymmetric} show the existence of asymmetric quantum product codes with specific parameters.
		
		\paragraph*{} Throughout the article, we denote by $q$ a power of a prime $p$. Also for ease of reference, most of the notations used in this survey are consistent with the various articles discussed.

	\section{Cyclic product codes}
	
	\paragraph*{} Cyclic product codes were first presented by Peterson in \cite{peterson1961error}, when he noticed that the product of two cyclic codes having generator polynomials $x + 1$ each, with lengths differing by one resulted in a cyclic code. Moreover, he showed that provided the lengths of the two subcodes were $n_1$ and $n_2$, the generator polynomial $g(x)$ for the product code was the following
	\begin{center}
		$g(x) = \frac{\left(x^{n_1} + 1\right)\left(x^{n_2} + 1\right)}{x + 1}$.
	\end{center}
	Later in \cite{burton1965cyclic} Burton and Weldon, were able to generalize this result to subcodes with arbitrary generator polynomials, and presented a weaker restriction on the lengths of these constituent codes.
	
	\paragraph*{} Algebraically, codewords of cyclic codes are represented as polynomials of at most certain degree, in particular one less than the length of the code. The authors enhanced this concept for two-dimensional product code with constituent subcodes as cyclic codes $[n_1, k_1]$ and $[n_2, k_2]$ having generator polynomials $g_1(x)$ and $g_2(x)$, respectively. Thus the polynomial representation of a codeword of a cyclic code of length $n_1n_2$ is the following
	\begin{center}
		$p(x) = p_0x^0 + p_1x^1 + \dots + p_{n_1n_2 - 1} x^{n_1n_2 - 1}$,
	\end{center}
	where in general the coefficients $p_i$ are linear combinations of the $k_1k_2$ information symbols.
	
	\paragraph*{} Consider a product code having  $n_1$ columns and $n_2$ rows and assume without loss of generality that $n_1 \geq n_2$. An ordered pair $(i, j)$ specifies each position in the array where $i$ and $j$ refer to the row and column, respectively.
	
	\paragraph*{} Next, the authors define a mapping $m(i, j)$ relating the $(i, j)$ element of the array to coefficient of the $(l + 1)$th term of the polynomial. i.e.,
	\begin{center}
		$p_l = a_{ij}$. 
	\end{center}
	where $l = m(i, j)$ and $a_{ij}$ is the $(i, j)$ element of the array.
	
	\paragraph*{} The following theorem from \cite{burton1965cyclic} provides the conditions on the array and a definition of $m(i, j)$ which results in the product of two cyclic codes to be cyclic:
	
	\begin{thm}[\textup{See Burton and Weldon \cite{burton1965cyclic}}]
		The product of two cyclic codes is itself a cyclic code provided
		\begin{enumerate}
			\item the lengths of the subcodes are relatively prime, i.e., $an_1 + bn_2 = 1$ for some integers $a$ and $b$ and,
			\item the mapping $m(i, j)$ is defined as
			\begin{center}
				$m(i, j) \equiv (j-i)bn_2 +i \, ^\ast, \, m(i, j) = 0, 1, \dots, n_1n_2-1$.
			\end{center}
		\end{enumerate}
		$^\ast$ All congruence relationships are modulo $n_1n_2$, and residue classes will be represented by the least positive integer in the class.
	\end{thm}

	\paragraph*{} Once we have a cyclic code the existence of a unique generator polynomial is assured. The following theorem from \cite{burton1965cyclic} provides the expression for the generator polynomial of a cyclic product code as a simple function of the generator polynomials of its subcodes.
	
	\begin{thm}[\textup{See Burton and Weldon \cite{burton1965cyclic}}]
		If $g_1(x)$ and $g_2(x)$ are the respective generator polynomials of the row and column subcodes of a cyclic product code, the generator polynomial of the product code is the greatest common divisor of
		\begin{center}
			$g_1(x^{bn_2}) g_2(x^{an_1})$ and $x^{n_1n_2} - 1.$
		\end{center}
	\end{thm}
	
	\paragraph*{} The generator polynomial of a cyclic product code composed of an arbitrary number of products can be found by successive applications of above the theorem.
	
	\paragraph*{} The product of two cyclic codes having relatively prime lengths is cyclic when the mapping given by Burton and Weldon is followed. Rajan, Madhusudhana and Siddiqi in \cite{rajan1993on}  generalized this notion for cyclic product codes with cyclic subcodes of equal length by following either a rowwise or columnwise mapping.
	
	\paragraph*{} The implementation of cyclic product codes in various communication systems and their error-correcting capabilities have been discussed extensively in the papers \cite{tang1966cyclic}, \cite{lin1970further} and \cite{bahl1971single-and}. Also, Lin and Weldon in \cite{lin1970further} demonstrated how the algebraic structure of cyclic product codes can be applied to establish the exact minimum distance of certain subclasses of BCH codes.

	\subsection{Bounds on the minimum distance of cyclic codes using cyclic product codes}

	\paragraph*{} The paper \cite{zeh2013generalizing} by Zeh et al. proposed a generalization of the Hartmann–Tzeng (HT) bound on the minimum distance of cyclic codes as a beautiful application of cyclic product codes. They embedded the given cyclic code into a cyclic product code to generalize the mentioned HT bound.
	
	\paragraph*{} Consider an $[n, k, d]$ cyclic code $C$ in the ring $\mathbb{F}_q[X]/(X^n - 1)$ generated by $g(X)$. The generator polynomial $g(X)$ has roots in the splitting field $\mathbb{F}_{q^s}$, where
	$n | (q^s - 1)$. Let $\alpha$ be a primitive element of order $n$ then the defining set $D_C$ of an $[n, k]$ cyclic code $C$ is:
	\begin{center}
		$D_C = \{0 \leq i \leq n - 1 \, | \, g(\alpha^i) = 0\}$.
	\end{center}
	
	\paragraph*{} Moreover, we follow the notations below for a given $z \in \mathbb{Z}$:
	\begin{align*}
		& D_C^{[z, \otimes]} \coloneqq \{(i \cdot z) \mod n \, | \, i \in D_C\},\\
		& D_C^{[z, +]} \coloneqq \{(i + z) \, | \, i \in D_C\}.
	\end{align*}

	\paragraph*{} The theorem below provides the polynomial form of the HT Bound.
	
	\begin{thm}\textup{(See Zeh et al. \cite{zeh2013generalizing}) (HT Bound)}
		Let an $[n, k]$ cyclic code $C$ with minimum distance $d$ be given. Let $\alpha$ denotes an element of order $n$. Let four integers $f, m, \delta$ and $\nu$ with $m \neq 0$ and $\gcd(n, m) = 1, \delta \geq 2$ and $\nu \geq 0$ be given, such that:
		\begin{center}
			$\sum \limits_{i = 0}^{\infty}c\left(\alpha^{f + im + j}\right)X^i \equiv 0 \mod X^{\delta - 1}, \quad \forall j = 0, \dots, \nu$,
		\end{center}
		holds for all $c(x) \in C$ and some integers $\delta \geq 2$ and $\nu \geq 0$.
		Then, $d \geq \delta + \nu$.
	\end{thm}
	
	Recalling the structure of cyclic product codes provided by Burton and Weldon in \cite{burton1965cyclic} we have the following theorem.
	
	\begin{thm}\textup{(See Zeh et al. \cite{zeh2013generalizing}) (Defining Set and Generator Polynomial of a Cyclic Product Code)}
		Let $A$ and $B$, respectively $[n_1, k_1]$ and $[n_2, k_2]$, be cyclic codes with defining sets $D_A$ and $D_B$ and generator polynomials $g_1(X)$ and $g_2(X)$. Let $an_1 + bn_2 = 1$ for some integers $a$ and $b$. Then, the generator polynomial $g(X)$ of the cyclic product code $A \otimes B$ is:
		\begin{center}
			$g(X) = \gcd \left(X^{n_1n_2} - 1, g_1\left(X^{bn_2}\right) \cdot g_2\left(X^{an_1}\right)\right)$.
		\end{center}
		Let $B_A = D_A^{[b, \otimes]}$ and let $A_B = D_B^{[a, \otimes]}$ as defined earlier. The
		defining set of the cyclic product code $C$ is:
		\begin{center}
			$D_C = \left\{\bigcup\limits_{i = 0}^{n_2 - 1}B_A^{[in_1, +]}\right\} \cup \left\{\bigcup\limits_{i = 0}^{n_1 - 1}A_B^{[in_2, +]}\right\}$.
		\end{center}
	\end{thm}

	The next theorem and proposition follows from Lin and Weldon's study of minimum distance of BCH codes in \cite{lin1970further}.

	\begin{thm}\textup{(See Zeh et al. \cite{zeh2013generalizing}) (BCH Bound Generalization)}
		Let an $[n_1, k_1]$ cyclic code $A$ with minimum distance $d_1$ and a second $[n_2, k_2]$ cyclic code $B$ with minimum distance $d_2$ and with $\gcd(n_1, n_2) = 1$ be given. Let $\alpha$ be an element of order $n_1$ in $\mathbb{F}_{q^{s_1}}$, $\beta$ of order $n_2$ in $\mathbb{F}_{q^{s_2}}$ respectively. Let the integers $f_1, f_2, m_1, m_2, \delta$ with $m_1 \neq 0, m_2 \neq 0, \gcd(n_1,m_1) = \gcd(n_2,m_2) = 1$ and $\delta \geq 2$ be given, such that:
		\begin{center}
			$\sum \limits_{i = 0}^{\infty}a\left(\alpha^{f_1 + im_1}\right) \cdot b\left(\beta^{f_2 + im_2}\right) X^i \equiv 0 \mod X^{\delta - 1}$
		\end{center}
		holds for all codewords $a(X) \in A$ and $b(X) \in B$. Then, we obtain:
		\begin{center}
			$d_1 \geq d^\ast = \ceil{\frac{\delta}{d_2}}$.
		\end{center}
	\label{thm : 2.5}
	\end{thm}

	\begin{prop}\textup{(See Zeh et al. \cite{zeh2013generalizing}) (BCH Bound of the Cyclic Product Code)}
		Let the integers $f_1, f_2, m_1 \neq 0, m_2 \neq 0$ and $\delta \geq 2$ and two cyclic codes $A$ and $B$ with $an_1 + bn_2 = 1$ be given as in Theorem $\ref{thm : 2.5}$
		Then, the two integers:
		\begin{align*}
			& f = f_1 \cdot b^2n_2 + f_2 \cdot a^2n_1 \quad \mathrm{and}\\
			& m = m_1 \cdot b^2n_2 + m_2 \cdot a^2n_1
		\end{align*}
		denote the parameters such that:
		\begin{center}
			$\sum \limits_{i = 0}^{\infty}c\left(\gamma^{f + im}\right) X^i \equiv 0 \mod X^{\delta - 1}$
		\end{center}
		 holds for all $c(X) \in A \otimes B$, where $\gamma$ is a primitive element of order $n_1n_2$ in $\mathbb{F}_{q^s}[X]$.
	\end{prop}

	Following are the two generalizations of the HT bound using cyclic product codes by Zeh et al.
	
	\begin{thm}\textup{(See Zeh et al. \cite{zeh2013generalizing}) (Generalized HT Bound I)}
		Let an $[n_1, k_1]$ cyclic code $A$ with minimum distance $d_1$ and a second $[n_2, k_2]$ cyclic code $B$ with minimum distance $d_2$ and with $\gcd(n_1, n_2) = 1$ be given. Let $\alpha$ be an element of order $n_1$ in $\mathbb{F}_{q^{s_1}}$, $\beta$ of order $n_2$ in $\mathbb{F}_{q^{s_2}}$ respectively. Let the integers $f_1, f_2, m_1, m_2, \delta$ and $\nu$ with $m_1 \neq 0, m_2 \neq 0, \gcd(n_1,m_1) = \gcd(n_2,m_2) = 1, \delta \geq 2$ and $\nu > 0$ be given, such that:
		\begin{center}
			$\sum \limits_{i = 0}^{\infty}a\left(\alpha^{f_1 + im_1 + j}\right) \cdot b\left(\beta^{f_2 + im_2 + j}\right) X^i \equiv 0 \mod X^{\delta - 1} \quad \forall j = 0, 1, \dots, \nu$
		\end{center}
		holds for all codewords $a(X) \in A$ and $b(X) \in B$. Then, the minimum distance $d_1$ of $A$ is lower bounded by:
		\begin{center}
			$d_1 \geq d^{\ast \ast} \coloneqq \ceil{\frac{\delta + \nu}{d_2}}$.
		\end{center}
	\end{thm}

	\begin{thm}\textup{(See Zeh et al. \cite{zeh2013generalizing}) (Generalized HT Bound II)}
		Let an $[n_1, k_1]$ cyclic code $A$ with minimum distance $d_1$ and a second $[n_2, k_2]$ cyclic code $B$ with minimum distance $d_2$ and with $\gcd(n_1, n_2) = 1$ be given. Let $\alpha$ be an element of order $n_1$ in $\mathbb{F}_{q^{s_1}}$, $\beta$ of order $n_2$ in $\mathbb{F}_{q^{s_2}}$ respectively. Let the integers $f_1, f_2, m_1, m_2, \delta$ and $\nu$ with $m_1 \neq 0, m_2 \neq 0, \gcd(n_1,m_1) = \gcd(n_2,m_2) = 1, \delta \geq 2$ and $\nu > 0$ be given, such that:
		\begin{center}
			$\sum \limits_{i = 0}^{\infty}a\left(\alpha^{f_1 + im_1 + j}\right) \cdot b\left(\beta^{f_2 + im_2}\right) X^i \equiv 0 \mod X^{\delta - 1} \quad \forall j = 0, 1, \dots, \nu$
		\end{center}
		holds for all codewords $a(X) \in A$ and $b(X) \in B$. Then, the minimum distance $d_1$ of $A$ is lower bounded by:
		\begin{center}
			$d_1 \geq d^{\ast \ast \ast} \coloneqq \ceil{\frac{\delta}{d_2} + \nu}$.
		\end{center}
	\end{thm}
	
	\section{Generalized Hamming weights of product codes}
	
	\paragraph*{} The notion of generalized Hamming weights for a product code in terms of those of its component codes was introduced by Wei and Yang in \cite{wei1993on}. They determined the weight hierarchy of a product code in terms of the weight hierarchies of its component codes, for several classes of
	direct product codes.
	
	\paragraph*{} We first briefly recall the concept of generalized Hamming weights. Let $C$ be an $[n, k]$ code. We recall the notion of generalized Hamming weights. The support of $C$, denoted by $\chi(C)$ is
	\begin{center}
		$\chi(C) := \{i : \exists \, (x_1, x_2, \dots x_n) \in C$ with $x_i \neq 0\}$.
	\end{center}
	For $1 \leq r \leq k$, the $r$th generalized Hamming weight of $C$ is defined as
	\begin{center}
		$d_r(C) := \min\{|\chi(D)| : D$ is a subcode of $C$ with $\dim(D) = r\}$ .	
	\end{center}
		
	\paragraph*{} Let $A \otimes B$ denote the product code of $[n_1, k_1]$ and $[n_2, k_2]$ codes $A$ and $B$ respectively i.e.
	\begin{center}
		$A \otimes B := \{x = [x_{ij}]_{1 \leq i \leq n_1, 1 \leq j \leq n_2} : (x_{1, j}, x_{2, j}, \dots, x_{n_1, j})  \in A \, \mathrm{for} \, \mathrm{each} \, j,$\medskip\\
		$\mathrm{and} \,$ $(x_{i, 1}, x_{i, 2}, \dots, x_{i, n_2}) \in B \, \mathrm{for} \, \mathrm{each} \, i\}$.
	\end{center}
	We define $(i, j) \in \chi(A \otimes B)$ provided there exists a codeword
	$x \in A \otimes B$ with $x_{i, j} \neq 0$. Following are some results on the generalized Hamming weights of product codes by Wei and Yang.
	
	\begin{thm}[\textup{See Wei and Yang \cite{wei1993on}}]
		Let $A$ and $B$ be two linear codes. Then
		\begin{enumerate}[(a)]
			\item $d_r(A \otimes B) \leq \min\{d_{r_1}(A)d_{r_2}(B) : r_1r_2 = r\}$,
			\item $d_r(A \otimes B) \leq \min\{d_{r_1}(A)d_{r_2}(B) - (r_1r_2 - r): r_1r_2 \geq r\}$,
			\item $d_1(A \otimes B) = d_1(A)d_1(B)$, and
			\item $d_2(A \otimes B) = \min\{d_1(A)d_2(B); d_2(A)d_1(B)\}$.
		\end{enumerate}
	\end{thm}

	\paragraph*{} A linear code $C$ is said to satisfy the chain condition if there exists subcodes $D_r$, for $1 \leq r \leq k$ such that rank$(D_r) = r, |\chi(D_r)| = d_r(C)$, and $D_{r-1}$ is a subcode of $D_r$. Some interesting results on chain condition follow.
	
	\begin{thm}[\textup{See Wei and Yang \cite{wei1993on}}]
		If $C$ satisfies the chain condition, so does its dual
		code $C^\perp$.
	\end{thm}
	
	\begin{thm}[\textup{See Wei and Yang \cite{wei1993on}}]
		The Hamming codes, dual Hamming codes, Reed-Muller codes of all orders, maximum-distance-separable codes, and extended Golay codes satisfy the chain condition.
	\end{thm}

	\paragraph*{} For $1 \leq r \leq k$, let $\Delta_r(C) = d_r(C) - d_{r - 1}(C)$. Let $A$ and $B$ denote $[n_A, k_A]$ and $[n_B, k_B]$ linear codes, respectively. Define
	\begin{align*}
		d_r^\ast(A \otimes B) := & \min\{\sum_{u = 1}^{k_A}\Delta_u(A)d_s(u)(B) :\\ 
		& \sum_{u = 1}^{k_A} s(u) = r, k_B \geq s(1) \geq s(2) \geq \dots \geq s(k_A) \geq 0\}.
	\end{align*}

	The main use of the concept of chain condition is relevant from the next theorem which provides relations between the weight hierarchy of a product code and the weight hierarchies of its component codes.

	\begin{thm}[\textup{See Wei and Yang \cite{wei1993on}}]
		If codes $A$ and $B$ both satisfy the chain condition, then
		\begin{enumerate}[(a)]
			\item $d_r^\ast(A \otimes B) = d_r^\ast(B \otimes A)$;
			\item $d_r(A \otimes B) \leq d_r^\ast(A \otimes B)$.
		\end{enumerate} 
	\end{thm}

	\section{Two-dimensional cyclic codes}
	
	\paragraph*{} Two-dimensional cyclic codes (or 2-D cyclic codes) were introduced by Ikai et al. in \cite{ikai1974two-dimensional}. Later Imai in \cite{imai1977a theory} gave a more algebraic viewpoint to it by defining a binary two-dimensional code as a set of $M \times N$ arrays (matrices) over $\mathbb{F}_2$.
	
	\paragraph*{} A two-dimensional code $C$ forms a subspace of the $MN$-dimensional vector space of the $M \times N$ arrays (matrices) over $\mathbb{F}_q$ then it is said to be linear. Again if we have a two-dimensional linear code such that for each code-array $C$, all the arrays obtained by permuting the columns or the rows of $C$ cyclically are also code-arrays then it is called a two-dimensional cyclic code.
	
	\paragraph*{} As we saw in the case of cyclic product codes, it is convenient to use a polynomial representation for $M \times N$ arrays. Let $P = [p_{i, j}]$ be an $M \times N$ array over $\mathbb{F}_q$. In the case of two-dimensional cyclic codes we represent the elements $p_{i, j}$ of $P$ as the coefficients of the bivariate polynomial
	\begin{center}
		$p(x,y) = \sum \limits_{(i, j) \in \Omega}^{} p_{i, j} x^iy^j$
	\end{center}
	where $\Omega = \{(i,j) \, | \, 0 \leq i < M, \, 0 \leq j < N; \, M, N \in\mathbb{Z}\}$. Instead of just using a single variable polynomial as in the case of product codes. As we will see using two variables help us to realize two-dimensional cyclic codes as ideals of quotient rings which is generally done in the case of cyclic codes.

	\paragraph*{} Denote by $\mathcal{P}$[$\Omega$] the set of polynomials of degree less than $M$ and $N$ with respect to $x$ and $y$ respectively over $\mathbb{F}_q$. Furthermore, for an arbitrary polynomial $f(x, y)$ over $\mathbb{F}_q$, let $\{f(x, y)\}_{\Omega}$ denote the polynomial in $\mathcal{P}[\Omega]$ such that
	\begin{center}
		$\{f(x, y)\}_{\Omega} \equiv f(x, y) \mod (x^M - 1, y^N - 1)$.
	\end{center}
	Thus we can write $\{f(x, y)\}_{\Omega}$ as
	\begin{center}
		$\{f(x, y)\}_{\Omega} = f(x, y) + a(x, y)(x^M - 1) + b(x, y)(y^N - 1)$
	\end{center}
	where $a(x, y)$ and $b(x,y)$ are polynomials over $\mathbb{F}_q$.
	
	\paragraph*{} The above polynomial representation helps us to define a two-dimensional cyclic code as a two-dimensional linear code such that for each codeword $c(x, y), \{xc(x, y)\}_\Omega$, and $\{yc(x,y)\}_\Omega$ are also codewords. Precisely saying a two-dimensional cyclic code is an ideal of the ring $\mathbb{F}_q[x,y]/\langle x^M - 1, y^N - 1\rangle$.
	
	\subsection{Relationship between quasi-cyclic codes and 2-D cyclic codes}
	
	\paragraph*{}
	
	A linear $[l.m,k,d]_{q}$ code $C$ of length $lm$, dimension $k$ and minimum distance $d$ over $\mathbb{F}_{q}$ is $l$-quasi-cyclic if every cyclic shift by $l$ of a codeword is again a codeword of $C$.
	
	\paragraph*{} Güneri and Özbudak presented a relationship between quasi-cyclic codes and 2-D cyclic codes in \cite{guneri2012a relation}. They considered a $q$-ary quasi-cyclic code $C$ of length $ml$ and index $l$, where both $m$ and $l$ are relatively prime to $q$ and showed that if the constituents of $C$ are cyclic codes, then $C$ can also be viewed as a 2-D cyclic code of size $m \times l$ over $\mathbb{F}_q$.
	
	\paragraph*{} They defined the ring $R \coloneqq \mathbb{F}_q[Y]/\langle Y^m - 1\rangle$. A linear code of length $l$ over $R$ is also an $R$-submodule of $R$. Take the map
	\begin{center}
		$\phi : \mathbb{F}_q^{ml} \rightarrow R^l,$\bigskip\\
		$c = \begin{pmatrix}
			c_{0, 0}, & \cdots, & c_{0, l - 1}\\
			c_{1, 0}, & \cdots, & c_{1, l - 1}\\
			& \vdots &\\
			c_{m - 1, 0}, & \cdots, & c_{m - 1, l - 1}
		\end{pmatrix} \mapsto \left(c_0(Y), c_1(Y), \dots, c_{l - 1}(Y)\right),$
	\end{center}
	where
	\begin{center}
		$c_j(Y) \coloneqq \sum\limits_{i = 0}^{m - 1}c_{ij}Y^i = c_{0, j} + c_{1, j}Y + c_{2, j}Y^2 + \cdots + c_{m - 1, j}Y^{m - 1} \in R$
	\end{center}
	for each $0 \leq j \leq l - 1$. In other words, an element of $R$ is constructed from each column of $c$. One can establish an one-to-one correspondence between index $l$ q.-c. codes of length $ml$ over $\mathbb{F}_q$ and
	linear codes of length $l$ over $R$ through the map $\phi$.
	
	\paragraph*{} $Y^m - 1$ is separable as $m$ is relatively prime to
	char$(\mathbb{F}_q)$. Assume
	\begin{center}
		$Y^m - 1 = f_1f_2 \dots f_r$
	\end{center}
	to be the unique factorization of $Y^m - 1$ into irreducible polynomials in $\mathbb{F}_q[Y]$. Rewrite $Y^m - 1$ as
	\begin{center}
		$Y^m - 1 = \delta g_1g_2 \dots g_sh_1h_1^\ast \dots h_th_t^\ast$,
	\end{center}
	where $g_i$'s are those $f_j$'s that are associate to their own reciprocals and the remaining $f_j$'s are grouped in pairs as $h_i, h_i^\ast$'s (so, $2t + s = r)$.
	
	\paragraph*{} Now, applying the Chinese Remainder Theorem we get
	\begin{center}
		$R = \bigg(\bigoplus_{i = 1}^{s} \mathbb{F}_q[Y]/(g_i)\bigg) \oplus \bigg(\bigoplus_{j = 1}^{t} \mathbb{F}_q[Y]/(h_j) \oplus \mathbb{F}_q[Y]/(h_j^\ast)\bigg).$
	\end{center}
	The following notations are used:
	\begin{center}
		$G_i = \mathbb{F}_q[Y]/(g_i), \quad H_j' = \mathbb{F}_q[Y]/(h_j), \quad H_j'' = \mathbb{F}_q[Y]/(h_j^\ast).$
	\end{center}
	Each of the above quotients are finite fields provided the fact that the polynomials are irreducible, (e.g. $[G_i : \mathbb{F}_q] = \deg g_i)$. Thus we have
	\begin{center}
		$R^l = \bigg(\bigoplus_{i = 1}^{s} G_i^l \bigg) \oplus \bigg(\bigoplus_{j = 1}^{t} (H_j')^l \oplus (H_j'')^l \bigg).$
	\end{center}
	So, we have the following decomposition of any linear code $C$ of length $l$ over $R$
	\begin{center}
		$C = \bigg(\bigoplus_{i = 1}^{s} C_i \bigg) \oplus \bigg(\bigoplus_{j = 1}^{t} C_j' \oplus C_j'' \bigg)$,
	\end{center}
	where $C_i, C_j', C_j''$ are linear codes of length $l$ over the fields $G_i, H_j', H_j''$ respectively. These length $l$ linear codes over various extensions of $\mathbb{F}_q$ are called the constituents of $C$.
	
	\begin{thm}\textup{(See Güneri and Özbudak \cite{guneri2012a relation})}
		Let $m$ be an integer such that $\gcd(m, q) = 1$ and let $\xi$ be a primitive $m$-th root of unity over $\mathbb{F}_q$. Assume that $Y^m - 1$ factors into irreducibles and adopt all the notation
		above. Let $U_i$, $V_j$, $W_j$ denote the $q$-cyclotomic cosets $\mod m$ corresponding to $G_i, H_j', H_j''$, respectively (i.e.
		corresponding to the powers of $\xi$ among the roots of $g_i, h_j, h_j^\ast$, respectively). Fix representatives $u_i \in U_i, v_j \in V_j, w_j \in W_j$ from cyclotomic cosets for each $i$ and $j$.\\
		Let $C_i$ over $G_i, C_j'$ over $H_j'$ and $C_j''$ over $H_j''$ be linear codes of length $l$ for all $i, j$. For codewords $\boldsymbol{\lambda_i} \in C_i, \boldsymbol{\beta_j} \in C_j', \boldsymbol{\gamma_j} \in C_j''$ and for each $0 \leq g \leq m - 1$, let
		\begin{align*}
			c_g = & c_g(\boldsymbol{\lambda_i}, \boldsymbol{\beta_j}, \boldsymbol{\gamma_j}) \coloneqq \sum\limits_{i = 1}^{s}\mathrm{Tr}_{G_i/\mathbb{F}_q}(\boldsymbol{\lambda_i} \xi^{-gu_i})\\
			& + \sum\limits_{j = 1}^{t}\left(\mathrm{Tr}_{H_j'/\mathbb{F}_q}\left(\boldsymbol{\beta_j} \xi^{-gv_j} \right) + \mathrm{Tr}_{H_j''/\mathbb{F}_q}\left(\boldsymbol{\gamma_j} \xi^{-gw_j} \right)\right),
		\end{align*}
		where the traces are applied to vectors coordinate wise so that $c_g$ is a vector of length $l$ over $\mathbb{F}_q$ for all $0 \leq g \leq m - 1$. Then the code
		\begin{center}
			$C = \left\{\left(c_0(\boldsymbol{\lambda_i}, \boldsymbol{\beta_j}, \boldsymbol{\gamma_j}), \dots, c_{m - 1}(\boldsymbol{\lambda_i}, \boldsymbol{\beta_j}, \boldsymbol{\gamma_j}\right) : \boldsymbol{\lambda_i} \in C_i, \boldsymbol{\beta_j} \in C_j', \boldsymbol{\gamma_j} \in C_j''\right\}$
		\end{center}
		is a q.-c. code over $\mathbb{F}_q$ of length $ml$ and index $l$.\\
		Conversely, every q.-c. code of length $ml$ and index $l$ is obtained through this construction.\label{thm : 3.1}
	\end{thm}
	
	\begin{thm}\textup{(See Güneri and Özbudak \cite{guneri2012a relation})}
		Let $n$ be relatively prime to $q$ and $\beta$ be a primitive $n$-th root of unity in some extension $\mathbb{E}$ of $\mathbb{F}_q$. Let $C$ be a $q$-ary cyclic code of length $n$ whose dual’s basic zero set is BZ$(C^\perp) = \{\beta^{i_1}, \dots, \beta^{i_s}\}$ (i.e. the generator polynomial of $C^\perp$ is $\prod_{j}m_j(x)$ where $m_j(x) \in \mathbb{F}_q[x]$ is the
		minimal polynomial of $\beta^{i_j}$ over $\mathbb{F}_q$). Then we have
		\begin{center}
			$C = \left\{\left(\mathrm{Tr}_{\mathbb{E}/\mathbb{F}_q}\left(c_1\beta^{ti_1} + \cdots + c_s\beta^{ti_s}\right)\right)_{0 \leq t \leq n - 1} : c_1, c_2, \dots, c_s \in E\right\}$.
		\end{center}
	\end{thm}
	
	\paragraph*{} Now, take q.-c. codes having nonzero constituents which are either full ambient spaces or proper cyclic codes. In particular, $m, l$ are both relatively prime to $q$ and $\xi, \nu$ are primitive $m$-th and $l$-th roots of unity,
	respectively, in some extension $\mathbb{F}$ of $\mathbb{F}_q$. When the nonzero constituent codes involved in Theorem \ref{thm : 3.1} are equal to their ambient spaces (i.e. they are full/maximal codes: $C_i = G_i^l, C_j' = (H_j')^l, C_j'' = (H_j'')^l$ for some $i, j$), then coordinates of the
	codewords will be arbitrary elements from those extensions of $\mathbb{F}_q$. In this case, all of the traces can be taken from a single extension of $\mathbb{F}_q$ as we see now. Note that having all constituents $C_i, C_j', C_j''$ equal to their ambient spaces would mean $C = R^l$ (equivalently, $C = \mathbb{F}_q^{lm}$), which is not interesting.
	
	\begin{thm}\textup{(See Güneri and Özbudak \cite{guneri2012a relation})}
		Suppose $C$ is a quasi cyclic code over $\mathbb{F}_q$ of length $ml$ and index $l$ whose nonzero constituents are full
		ambient spaces. Namely, let
		\begin{center}
			$C = \left(G_{i_1}^l \oplus \cdots \oplus G_{i_e}^l\right) \oplus \left(\left(H_{j_1}'\right)^l \oplus \cdots \oplus \left(H_{j_f}'\right)^l\right) \oplus \left(\left(H_{k_1}''\right)^l \oplus \cdots \oplus \left(H_{k_h}''\right)^l\right)$,
		\end{center}
		where $\{i_1, i_2 \dots, i_e\} \subset \{1, 2, \dots, s\}$ and $\{j_1, j_2, \dots, j_f\}, \{k_1, k_2, \dots, k_h\} \subset \{1, 2, \dots, t\}$. Let $u_{i_1}, u_{i_2}, \dots, u_{i_e}, \, v_{j_1}, v_{j_2} \dots, v_{j_f}, \, w_{k_1}, \dots, w_{k_h}$ denote fixed representatives of $q$-cyclotomic cosets $\mod m$ corresponding
		to the fields appearing in above equation. Then
		\begin{center}
			$\dim(C) = l\dim(D)$ and $d(C) = ld(D)$,
		\end{center}
		where $D$ is the $q$-ary cyclic code of length $m$ whose dual’s basic zero set is
		\begin{center}
			BZ$(D^\perp) = \left\{\xi^{-u_{i_1}}, \dots, \xi^{-u_{i_e}}, \, \xi^{-v_{j_1}}, \dots, \xi^{-v_{j_f}}, \, \xi^{-w_{k_1}}, \dots, \xi^{-w_{k_h}}\right\}$.
		\end{center}
	\end{thm}

	\paragraph*{} Consider quasi cyclic code $C$ with cyclic constituents. Both $m$ and $l$ are relatively prime to $q$ and $\xi, \nu$ are primitive $m$-th and $l$-th roots of unity, respectively, in the extension $\mathbb{F}$ of $\mathbb{F}_q$. Take the q.-c. code $C$ of length $ml$, index $l$ over $\mathbb{F}_q$ such that
	\begin{center}
		$C = (C_1 \oplus \cdots \oplus C_e) \oplus (D_1 \oplus \cdots \oplus D_f) \oplus (E_1 \oplus \cdots \oplus E_h)$,
	\end{center}
	where $C_a \subset G_{i_a}^l \, (1 \leq a \leq e), \, D_b \subset (H_{j_b}')^l \, (1 \leq b \leq f), \, D_b \subset (H_{k_d}'')^l \, (1 \leq d \leq h)$ are all cyclic codes of length $l$ over various extensions of $\mathbb{F}_q$. These cyclic codes can be defined by their duals’ basic zero sets as follows:\bigskip\\
	$\begin{matrix}
		 BZ(C^\perp_1) = \left\{\eta^{x_{1, 1}} \dots, \eta^{x_{1, k_1}}\right\} & BZ(D^\perp_1) = \left\{\eta^{y_{1, 1}} \dots, \eta^{y_{1, m_1}}\right\} & BZ(E^\perp_1) = \left\{\eta^{z_{1, 1}} \dots, \eta^{z_{1, n_1}}\right\}\\
		 \vdots & \vdots & \vdots\\
		 BZ(C^\perp_e) = \left\{\eta^{x_{e, 1}} \dots, \eta^{x_{e, k_e}}\right\} &  BZ(D^\perp_f) = \left\{\eta^{y_{f, 1}} \dots, \eta^{y_{f, m_f}}\right\} &  BZ(E^\perp_h) = \left\{\eta^{z_{h, 1}} \dots, \eta^{z_{h, n_h}}\right\}.
	\end{matrix}$
	
	\begin{thm}\textup{(See Güneri and Özbudak \cite{guneri2012a relation})}
		Let $m$ and $l$ be integers which are relatively prime to $q$. Let C be a quasi cyclic code over $\mathbb{F}_q$ of length $ml$ and index $l$ such that
		\begin{center}
			$C = (C_1 \oplus \cdots \oplus C_e) \oplus (D_1 \oplus \cdots \oplus D_f) \oplus (E_1 \oplus \cdots \oplus E_h)$,
		\end{center}
		where each constituent is a cyclic code of length $l$ over various extensions of $\mathbb{F}_q$. Assume that the cyclic constituents are defined by their duals’ basic zero sets as in Theorem $4.3$. Then $C$ is a $2$-D cyclic code of size $m \times l$ over $\mathbb{F}_q$ whose dual’s basic zero set is
		\begin{center}
			$BZ(C^\perp) = S_1 \cup \dots \cup S_e \cup S_1' \cup \dots \cup S_f' \cup S_1'' \cup \dots \cup S_h''$
		\end{center}
	  with the notation of Lemma $3.4.$ from \textup{\cite{guneri2012a relation}}.
	\end{thm}

	\subsection{Generator matrix for two-dimensional cyclic code of arbitrary length}
	
	\paragraph*{} At the beginning of this section we saw that two-dimensional cyclic codes of length $n = sl$ over the finite field $\mathbb{F}_q$ are ideals
	of the polynomial ring $\mathbb{F}_q[x, y]/\langle x^s-1, y^l-1\rangle$. The paper \cite{sepasdar2017generator} by Sepasdar provides a method to find generator matrix for these two-dimensional cyclic codes of arbitrary length $n = sl$ over the finite field $\mathbb{F}_q$. A new method to characterize ideals of the ring $\mathbb{F}_q[x, y]/\langle x^s-1, y^l-1\rangle$ corresponding to two-dimensional cyclic codes is presented, and generator polynomials are found for these ideals. Finally, these polynomials are used to obtain the generator matrix for corresponding two-dimensional cyclic codes.
	
	\paragraph*{} Let $R \coloneqq \mathbb{F}_q[x, y]/\langle x^s-1, y^l-1\rangle$ and $S \coloneqq  \mathbb{F}_q[x]/\langle x^s-1\rangle$. Suppose that
	$I$ is an ideal of $R$. Then we can construct ideals $I_i$ of $S \, (i = 0, \dots, l - 1)$ and show that the monic generator polynomials of $I_i$ form a generating set for $I$ as
	\begin{center}
		$\mathbb{F}_q[x, y]/\langle x^s-1, y^l-1\rangle \cong \left(\mathbb{F}_q[x]/\langle x^s-1 \rangle \right) [y]/\langle y^l-1 \rangle$,
	\end{center}
	an arbitrary element $f(x, y) \in I$ can be written uniquely as $f(x, y) = \sum_{i = 0}^{l - 1} f_i(x)y^i$, where $f_i(x) \in S$ for $i = 0, \dots, l - 1$. Take
	\begin{center}
		$I_0 = \{g_0(x) \in S : \,  \exists \, g(x, y) \in I$ such that $g(x, y) = \sum_{i = 0}^{l - 1} g_i(x)y^i\}$.
	\end{center}
	It can be shown that $I_0$ is an ideal of the principal ideal ring $S$. Thus, $\exists$ a unique monic polynomial $p_0^0(x) \in S$ such that $I_0 = \langle p_0^0(x) \rangle$ and $p_0^0(x)$  is a divisor of $x^s - 1$. So $\exists \, p'_0(x)$ in $\mathbb{F}_q$ such that $x^s - 1 = p'_0(x)p^0_0(x)$. Now, consider the following equations
	\begin{align*}
		f(x, y) & = f_0(x) + f_1(x)y + \cdots + f_{l - 1}(x)y^{l - 1} \\
		yf(x, y) & = f_0(x)y + f_1(x)y^2 + \cdots + f_{l - 1}(x)y^l \\
		& = f_{l - 1}(x) + f_0(x)y + f_1(x)y^2 + \cdots + f_{l - 2}(x)y^{l - 1}. \hspace{1cm} (y^l = 1 \, \mathrm{in} \, R)
	\end{align*}
	Since $I$ is an ideal of $R$, we have $yf(x, y) \in I$. So by definition of $I_0, \, f_{l - 1}(x) \in
	I_0$. Similarly, one can prove that $f_i(x) \in I_0 = \langle p_0^0(x) \rangle$ for
	$i = 1, \dots, l - 2$. So
	\begin{center}
		$f_i(x) = p^0_0(x) q_i(x)$
	\end{center}
	for some $q_i(x) \in S$. Now, $p_0^0(x) \in I_0$ so by the definition of $I_0$, $\exists \, \mathbf{p}_0(x, y) \in I$ such that
	\begin{center}
		$\mathbf{p}_0(x, y) = \sum_{i = 0}^{l - 1} p_i^0(x) y^i$.
	\end{center}
	Again since $I$ is an ideal of $R$, $y^i\mathbf{p}_0(x, y) \in I$ for $i = 1, \dots, l - 1$. So by the definition of $I_0, p^0_i(x) \in I_0 = \langle p_0^0(x) \rangle$. Therefore,
	\begin{center}
		$p_i^0(x) = p_0^0(x)t_i^0(x)$
	\end{center}
	for some $t_i^0(x) \in S$, and so
	\begin{center}
		$\mathbf{p}_0(x, y) = p_0^0(x) + \sum_{i = 1}^{l - 1} p_0^0(x) t_i^0(x) y^i$.
	\end{center}
	Set
	\begin{center}
		$h_1(x, y) \coloneqq f(x, y) - \mathbf{p}_0(x, y)q_0(x) = \sum_{i = 1}^{l - 1} f_i(x) y^i - q_0(x) \sum_{i = 1}^{l - 1} p_i^0(x) y^i$.
	\end{center}

	 \paragraph*{} Since $f(x, y)$ and $\mathbf{p}_0(x, y)$ are in $I$ which is an ideal of $R$, $h_1(x, y)$ is a polynomial in $I$. Also $h_1(x, y)$ is in the form of $h_1(x, y) = \sum_{i = 1}^{l - 1} h_i^1(x) y^i$ for some $h_1^i(x) \in S$. Now, take
	 \begin{center}
	 	$I_1 = \{g_1(x) \in S :\, \exists \, g(x, y) \in I$ such that $g(x, y) = \sum_{i = 1}^{l - 1} g_i(x) y^i\}$.
	 \end{center}
	By similar method being applied for $I_0$, we can prove that $I_1$ is an ideal of $S$. Thus applying the same procedure we obtain the polynomials
	\begin{center}
		$h_2(x, y), \dots, h_{l - 1}(x, y), \mathbf{p}_1(x, y), \dots, \mathbf{p}_{l - 1}(x, y)$
	\end{center}
	in $I$ and polynomials $q_2(x), \dots, q_{l - 1}(x)$ in $S$ such that for an arbitrary element $f(x, y) \in I$ we have
	\begin{align*}
		h_1(x, y) & \coloneqq f(x, y) - \mathbf{p}_0(x, y)q_0(x)\\
		h_2(x, y) & \coloneqq h_1(x, y) - \mathbf{p}_1(x, y)q_1(x)\\
		h_3(x, y) & \coloneqq h_2(x, y) - \mathbf{p}_2(x, y)q_1(x)\\
		& \vdots\\
		h_{l - 1}(x, y) & \coloneqq h_{l - 2}(x, y) - \mathbf{p}_{l - 2}(x, y)q_{l - 2}(x)\\
		h_{l - 1}(x, y) & = q_{l - 1}(x) \mathbf{p}_{l - 1}(x, y).
	\end{align*}
	So
	\begin{center}
		$f(x, y) = \mathbf{p}_0(x, y)q_0(x) + \mathbf{p}_1(x, y)q_1(x) + \cdots + \mathbf{p}_{l - 1}(x, y)q_{l - 1}(x)$.
	\end{center}
	Since $\mathbf{p}_i(x, y) \in I$ for $i = 0, \dots, l - 1$ and $f(x, y)$ is an arbitrary element of $I$ and $I$ is an ideal of $R$, we have
	\begin{center}
		$I = \langle \mathbf{p}_0(x, y), \dots, \mathbf{p}_{l - 1}(x, y) \rangle$,
	\end{center}
	 where $\mathbf{p}_j(x, y) = \sum_{i = j}^{l - 1} p_0^0(x) t^j_i y^i$, for some $t^j_i \in S$. So $\{\mathbf{p}_0(x, y), \mathbf{p}_1(x, y) \dots, \mathbf{p}_{l - 1}(x, y)\}$ is a set
	 of generating polynomials for I.
	 
	 \paragraph*{} From the above discussion, we get the following theorem that provides the generator matrix for two-dimensional cyclic codes.
	 
	\begin{thm}\textup{(See Sepasdar \cite{sepasdar2017generator})}
		Suppose that $I$ is an ideal $\mathbb{F}_q[x, y]/\langle x^s-1, y^l-1\rangle$ and is generated by $\{\mathbf{p}_0(x, y), \dots, \mathbf{p}_{l - 1}(x, y)\}$, which is obtained from the above method. Then the set
		\begin{align*}
			& \{\mathbf{p}_0(x, y), x\mathbf{p}_0(x, y), \dots, x^{s - a_0 - 1}\mathbf{p}_0(x, y),\medskip\\
			& \mathbf{p}_1(x, y), x\mathbf{p}_1(x, y), \dots, x^{s - a_1 - 1}\mathbf{p}_1(x, y),\medskip\\
			& \hspace{3.5cm} \vdots\medskip\\
			& \mathbf{p}_{l - 1}(x, y), x\mathbf{p}_{l - 1}(x, y), \dots, x^{s - a_{l - 1} - 1}\mathbf{p}_{l - 1}(x, y)\}
		\end{align*}
		forms an $\mathbb{F}_q$-basis for $I$, where $a_i = \deg(p^i_i(x))$.
	\end{thm}

	\subsection{Two-dimensional cyclic codes as ideals of $\mathbb{F}_q[x, y]/\langle x^s - 1, y^{2^k} - 1 \rangle$}

	\paragraph*{} Sepasdar and Khashyarmanesh in \cite{sepasdar2016characterizations} characterized some two-dimensional cyclic codes (abbreviated as TDC codes in this section) corresponding to the ideals of $\mathbb{F}_q[x, y]/\langle x^s - 1, y^{2^k} - 1 \rangle$. Some of the relevant results from their paper in context of this survey are stated here.
	
	\begin{thm}\textup{(See Sepasdar and Khashyarmanesh \cite{sepasdar2016characterizations})}
		Let $I$ be an ideal of $\mathbb{F}_q[x, y]/\langle x^s - 1, y^{2^k} - 1 \rangle$ with the generators $p_1(x)(1 + y)$ and $p_2(x)(1 - y)$ for some monic polynomials $p_1(x)$ and $p_2(x)$ in $\mathbb{F}_q[x]/\langle x^s - 1 \rangle$ with $\deg(p_1(x)) = a$ and $\deg(p_2(x)) = b$. Then every element of $I$ has the form
		\begin{center}
			$a_1(x)p_1(x)(1 + y) + a_2(x)p_2(x)(1 - y)$,
		\end{center}
		where $a_1(x)$ and $a_2(x)$ are polynomials in the ring $\mathbb{F}_q[x]/\langle x^s - 1 \rangle$ with $\deg(a_1(x)) < s - a$ and $\deg(a_2(x)) < s - b$.
	\end{thm}
	
	\begin{thm}\textup{(See Sepasdar and Khashyarmanesh \cite{sepasdar2016characterizations})}
		The number of TDC codes of length $n =2s$ with exactly $2$ generators is
		\begin{center}
			$\left(\left(p^l + 1\right)^r\right)
			\left(\left(p^l + 1\right)^r - 1\right)$,
		\end{center}
		where $x^s - 1 = \prod_{i = 1}^{r}p_i(x)^{p^l}$ for some distinct monic irreducible polynomials $p_i(x)$ for $i = 1, 2, \dots, r$.
	\end{thm}
	
	\begin{thm}\textup{(See Sepasdar and Khashyarmanesh \cite{sepasdar2016characterizations})}
		Suppose that $C$ is a TDC code of length $n = 2s$ with generator polynomials $p_1(x)(1 + y)$ and $p_2(x)(1 - y)$. Also assume that $\deg(p_1(x)) = a$ and $\deg(p_2(x)) = b$. Then
		\begin{center}
			$G = 
			\begin{pmatrix}
				p_1(x)(1 + y)\\
				xp_1(x)(1 + y)\\
				\vdots\\
				x^{s - a - 1}p_1(x)(1 + y)\\
				p_2(x)(1 - y)\\
				xp_2(x)(1 - y)\\
				\vdots\\
				x^{s - b - 1}p_2(x)(1 - y)
			\end{pmatrix}$
		\end{center}
		is a generator matrix of $C$.
	\end{thm}

	\begin{prop}\textup{(See Sepasdar and Khashyarmanesh \cite{sepasdar2016characterizations})}
		The dual code of a TDC code is also a TDC code.
	\end{prop}

	\paragraph*{} For a polynomial $f(x)$ with $\deg(f(x)) = k$ we define its reciprocal polynomial denoted by $f_R(x)$ as $x^kf(1/x)$. Consider $h'(x) = \sum_{i = 0}^{s - 1}h'_ix^i$ and $h(x) = \sum_{i = 0}^{s - 1}h_ix^i$. Assume $\mathbf{h_i}$ to be the vector obtained from $(h_0, \dots, h_{s - 1})$ by a cyclic shift $i$ positions. Then we have the following	\begin{center}
		$h'(x)h(x) = x^s - 1$
	\end{center}
	which implies that $(h'_{s - 1}, \dots, h'_0) \cdot \mathbf{h_i} = 0$, for $i =0, \dots, s - 1$, and thus $h'_R(x)$ is a codeword in $C^\perp$.
	
	\begin{thm}\textup{(See Sepasdar and Khashyarmanesh \cite{sepasdar2016characterizations})}
		If $C$ is a TDC code of length $n = 2s$ with generator polynomials $p_1(x)(1 + y)$ and $p_2(x)(1 - y)$ such that $\deg(p_1(x)) = a$, $p_1'(x)p_1(x) = x^s - 1, \deg(p_2(x)) = b$ and $p_2'(x)p_2(x) = x^s - 1$. Then
		\begin{center}
			$H = 
			\begin{pmatrix}
				(p_1'(x))_R(1 + y)\\
				x(p_1'(x))_R(1 + y)\\
				\vdots\\
				x^{a - 1}(p_1'(x))_R(1 + y)\\
				(p_2'(x))_R(1 - y)\\
				x(p_2'(x))_R(1 - y)\\
				\vdots\\
				x^{b - 1}(p_2'(x))_R(1 - y)
			\end{pmatrix}$
		\end{center}
		is the generator matrix of $C^\perp$.
	\end{thm}
	
	\begin{thm}\textup{(See Sepasdar and Khashyarmanesh \cite{sepasdar2016characterizations})}
		Suppose that $C$ is a TDC code of length $n = 4s$ and it’s generating set of polynomials is
		\begin{center}
			$\{p_1(x)(1 + y + y^2 + y^3), p_2(x)(1 - y + y^2 - y^3), p_3(x)(1 - y^2)\}$
		\end{center}
		with $\deg(p_i(x)) = a_i$ for $i = 1, 2, 3$. Then
		\begin{align*}
			\{ & p_1(x)(1 + y + y^2 + y^3), xp_1(x)(1 + y + y^2 + y^3), \dots, x^{s - a_1 - 1}p_1(x)(1 + y + y^2 + y^3),\\
			& p_2(x)(1 - y + y^2 - y^3), xp_2(x)(1 - y + y^2 - y^3), \dots, x^{s - a_2 - 1}p_1(x)(1 - y + y^2 - y^3),\\
			& p_3(x)(1 - y^2), xp_3(x)(1 - y^2), \dots, x^{s - a_3 - 1}p_3(x)(1 - y^2),\\
			& p_3(x)y(1 - y^2), xp_3(x)y(1 - y^2), \dots, x^{s - a_3 - 1}p_3(x)y(1 - y^2),\},
		\end{align*}
		is an independent set, and so the elements of this set form the rows of a generator matrix of the code $C$.
	\end{thm}
	
	\begin{thm}\textup{(See Sepasdar and Khashyarmanesh \cite{sepasdar2016characterizations})}
		Suppose that $C$ is a TDC code of length $n = 4s$ with the generator set
		\begin{center}
			$\{p_1(x)(1 + y + y^2 + y^3), p_2(x)(1 - y + y^2 - y^3), p_3(x)(1 - y^2)\}$
		\end{center}
		such that $\deg(p_i(x)) = a_i$ for $i = 1, 2, 3$. Let $p_i'(x)$ be a polynomial such that $p_i'(x)p_i(x) = x^s - 1$, for $i = 1, 2, 3$. Then
		\begin{align*}
			D = \{ & (p_1'(x))_R(1 + y)(1 + y^2), x(p_1'(x))_R(1 + y)(1 + y^2), \dots, x^{a_1 - 1}(p_1'(x))_R(1 + y)(1 + y^2),\\
			& (p_2'(x))_R(1 - y)(1 + y^2), x(p_1'(x))_R(1 - y)(1 + y^2), \dots, x^{a_2 - 1}(p_2'(x))_R(1 - y)(1 + y^2),\\
			& (p_3'(x))_R(1 - y^2), x(p_3'(x))_R(1 - y^2), \dots, x^{a_3 - 1}(p_3'(x))_R(1 - y^2),\\
			& (p_3'(x))_R \, y(1 - y^2), x(p_3'(x))_R \, y(1 - y^2), \dots, x^{a_3 - 1}(p_3'(x))_R \, y(1 - y^2),\},
		\end{align*}
		is an independent set, and so the polynomials of $D$ form the rows of a generator matrix of the code $C^\perp$.
	\end{thm}	
	
	\begin{thm}\textup{(See Sepasdar and Khashyarmanesh \cite{sepasdar2016characterizations})}
		Suppose that $C$ is a TDC code of length $n = 2^ks$. Then the corresponding ideal $I$ has the generating set of polynomials as follows:
		\begin{align*}
			I = \Bigg\langle & p_1(x)\left(\sum_{i = 0}^{2^k - 1} y^i\right), p_2(x)\left(\sum_{i = 0}^{2^k - 1} (-1)^iy^i\right), p_3(x)\left(\sum_{i = 0}^{2^{k - 1} - 1} (-1)^iy^{2i}\right),\\
			& p_4(x)\left(\sum_{i = 0}^{2^{k - 2} - 1} (-1)^iy^{4i}\right), \dots, p_{k+1}(x)\left(\sum_{i = 0}^{2^l - 1} (-1)^iy^{2^{k - 1}i}\right) \Bigg\rangle.
		\end{align*}
	\end{thm}
	
	The authors obtained generating set of polynomials for ideals corresponding to TDC codes of length $n = s2^k$ for each integer value of $s$ and $k$. But it was not extended for arbitrary TDC codes. For example, the method does not work for TDC codes over the ring $\mathbb{F}_q[x, y]/\langle x^3 - 1, y^3 - 1\rangle$.

	\subsection{Relation between weights of codewords and Artin–Schreier curves}
		
	\paragraph*{} Algebraic curves over finite fields have been used for the weight computations of certain cyclic codes. In \cite{guneri2004artin}, Güneri applied the above to 2-D cyclic codes by introducing a trace representation for such codes. We provide a brief summary of the results presented.
	
	\paragraph*{} Let $\mathbb{F}_{q^m}$ is the degree $m$ extension of $\mathbb{F}_q$ where $m > 1$. The equation $y^q - y = f(x)$  is referred as A–S (Artin–Schreier) curve, where $f(x) \in \mathbb{F}_{q^m}[x]$. The number of affine $\mathbb{F}_{q^m}$ -rational points on the curve is exactly the number of solutions $(x, y)$ to this equation in $\mathbb{F}_{q^m} \times \mathbb{F}_{q^m}$. In \cite{guneri2002a bound}, the family of Artin–Schreier curves $\mathcal{F} = \{y^q - y = \lambda_1x^{i_1} + \lambda_2x^{i_2} + \dots + \lambda_sx^{i_s}; \lambda_1, \dots, \lambda_s \in \mathbb{F}_{q^m}\}$ with positive integers $i_1, \dots i_s$ is considered. 
	
	\begin{thm}\textup{(See Güneri \cite{guneri2004artin})}
	 	Let $X : y^q - y = \lambda_1x^{i_1} + \lambda_2x^{i_2} + \dots + \lambda_sx^{i_s}$, where $\lambda_j$'s are in $\mathbb{F}_{q^m}$ and suppose $\#_{\mathbb{F}_{q^m}}(X)$ denote the number of affine $\mathbb{F}_{q^m}$- rational points of $X$. Let $B_j$ denote the $q$-cyclotomic coset containing $i_j \mod q^m - 1$, for all $j$, and assume that $B_j$'s
		are pairwise disjoint. Then the following are equivalent:
		\begin{enumerate}
			\item $\#_{\mathbb{F}_{q^m}}(X) = q^{m+1}$,
			\item $|B_j| = \delta_j \neq m$ and $\textup{tr}_{q^mq^{\delta_j}}(\lambda_j) = 0$; for all $j = 1, 2, \dots, s$.
		\end{enumerate}
	\end{thm}

	Let $\alpha_1$ be a primitive $n_1$th root of unity and $\alpha_2$ be a primitive
	$n_2$th root of unity, where $n_1$ and $n_2$ to be relatively prime to $p$. Taking both of these elements in the smallest extension $\mathbb{F}_{q^s}$ of $\mathbb{F}_q$ such that $n_1$ and $n_2$ divide $q^s - 1$ we consider the following set
	\begin{center}
		$\Omega = \{(\alpha_1^i, \alpha_2^j); 0 \leq i \leq n_1 - 1, 0 \leq j \leq n_2 - 1\}$.
	\end{center}
	Define the $\mathbb{F}_q$-conjugacy class containing $(\alpha_1^i, \alpha_2^j)$ to be
	\begin{center}
		$S = [(\alpha_1^i, \alpha_2^j)] =\{(\alpha_1^i, \alpha_2^j), (\alpha_1^{iq}, \alpha_2^{jq}), \dots, (\alpha_1^{iq^{\delta -1}}, \alpha_2^{jq^{\delta -1}})\}$,
	\end{center}
	where $\delta$ is the LCM of the degrees $\alpha_1^i$ and $\alpha_2^j$ over $\mathbb{F}_q$.
	
	\paragraph*{} The letter $U$ is used only for either a single class or a finite union of $\mathbb{F}_q$-conjugacy classes in $\Omega$.
	
	\paragraph*{} For $U \subset \Omega$, the ideal corresponding to $U$ is defined as
	\begin{center}
		$I(U) = \{f(x, y) \in \mathbb{F}_q[x, y]; f(a) = 0, \forall a \in U\}$.
	\end{center}
	 
	 Note that $x^{n_1} - 1$ and $y^{n_2} - 1$ are in $I(U)$ for any $U \subset \Omega$. Therefore, $I(U)/(x^{n_1} - 1, y^{n_2} - 1) \subset \mathbb{F}_q[x, y]/(x^{n_1} - 1, y^{n_2} - 1)$ is a 2-D cyclic code. Then
	 \begin{center}
	 	$Z(\tilde{J}) = \{(\gamma, \beta) = 0, \forall f \in J\}$
	 \end{center}
 	 is called the zero set of the 2-D cyclic code $\tilde{J}$. Note that since $x^{n_1} - 1$ and $y^{n_2} - 1$ are in $J$, the zero set $Z(\tilde{J})$ is a subset of $\Omega$ and it is either a single $\mathbb{F}_q$-conjugacy class or a finite union of $\mathbb{F}_q$-conjugacy classes.
	 
	\begin{thm}\textup{(See Güneri \cite{guneri2004artin})}
		Let $U$ be a subset of $\Omega$ and let $\overline{U}$ denote $\Omega - \overline{U}$ Consider the $2$-D cyclic code $C_U = \tilde{I}(U) = I(U)/(x^{n_1} - 1, y^{n_2} - 1)$ corresponding to $U$. The dimension of $C_U$
		is given by
		\begin{center}
			$\dim_{\mathbb{F}_q} C(U) = |\overline{U}|$.
		\end{center}
	\end{thm}

	\begin{prop}\textup{(See Güneri \cite{guneri2004artin})}
		For the $2$-D cyclic code $C_U = \tilde{I}(U) = I(U)/(x^{n_1} - 1, y^{n_2} - 1)$, its
		dual code is the $2$-D cyclic code $C_{\overline{U}^{-1}} = \tilde{I}(\overline{U}^{-1}) = I(\overline{U}^{-1})/(x^{n_1} - 1, y^{n_2} - 1)$; which has the zero set
		\begin{center}
			$Z(C_U^\perp) = Z(C_{\overline{U}^{-1}}) = \bar{U}^{-1} = \Omega - U^{-1}$,
		\end{center}
		where
		\begin{center}
			$U^{-1} = \{(\mu_1^{-1}, \mu_2^{-1}); (\mu_1, \mu_2) \in U\}$.
		\end{center}
	\end{prop}

	\begin{coro}\textup{(See Güneri \cite{guneri2004artin})}
		The dimension of a $2$-D cyclic code is equal to the number of zeros of its
		dual code.
	\end{coro}
	
	\paragraph*{} If the zero set of a 2-D cyclic code $C$ is the union of the $\mathbb{F}_q$-conjugacy
	classes $S_\gamma = [(\alpha_1^{i_\gamma}, \alpha_2^{j_\gamma})]$, where $\gamma$ is in some index set $\mathcal{I}$; then the set
	\begin{center}
		$\{(\alpha_1^{i_\gamma}, \alpha_2^{j_\gamma}); \gamma \in \mathcal{I}\}$
	\end{center}
	is called a basic zero set of $C$ and denoted BZ$(C)$. Similarly, a basic nonzero set of $C$ can be defined and denote it by BNZ$(C)$.
	
	\paragraph*{} Consider
	\begin{center}
		$\Omega = \{(\alpha^i, \alpha^j); 0 \leq i, j \leq q^m - 2\} = \mathbb{F}_{q^m}^\ast \times \mathbb{F}_{q^m}^\ast$,\medskip\\
		$U = [(\alpha^{i_1}, \alpha^{j_1})] \cup [(\alpha^{i_2}, \alpha^{j_2})] \cup \cdots \cup [(\alpha^{i_s}, \alpha^{j_s})]$,
	\end{center}
	where $i_\gamma, j_\gamma$ are in the set $\{1, 2, \dots, q^m - 2\}$ and $[(\alpha^{i_\gamma}, \alpha^{j_\gamma})] = [(\alpha^{i_\gamma'}, \alpha^{j_\gamma'})]$ if $\gamma = \gamma'$, i.e. the
	pairs $(\alpha^{i_\gamma}, \alpha^{j_\gamma})$, for $\gamma = 1, 2, \dots, s$, are representatives of distinct $\mathbb{F}_q$-conjugacy classes.
	
	\paragraph*{} Define $C$ to be the 2-D cyclic code of area $(q^m - 1) \times (q^m - 1)$ over $\mathbb{F}_q$ which
	has the following zero set:
	\begin{center}
		$Z(C) = \Omega - U^{-1} = \overline{U}^{-1}$.
	\end{center}
	Let $D'$ be the 2-D cyclic code of area $(q^m - 1) \times (q^m - 1)$ defined
	over $\mathbb{F}_{q^m}$ by the zero set
	\begin{center}
		$Z(D') = \{(\alpha^{i_1}, \alpha^{j_1}), (\alpha^{i_2}, \alpha^{j_2}), \dots, (\alpha^{i_s}, \alpha^{j_s})\}$.
	\end{center}

	\paragraph*{} Denote all the elements of the multiplicative group $\mathbb{F}^\ast_{q^m}$ by $A$ i.e. for a primitive element $\alpha$ of $\mathbb{F}_{q^m}$ we have,
	\begin{center}
		$A = \mathbb{F}^\ast_{q^m} = \{\alpha^0, \alpha^1, \alpha^2, \dots, \alpha^{q^m - 2}\}$
	\end{center}

	\begin{thm}\textup{(See Güneri \cite{guneri2004artin})}
		Let $C$ be the $2$-D cyclic code of area $(q^m - 1) \times (q^m - 1)$ over $\mathbb{F}_q$ whose
		dual’s zero set is
		\begin{center}
			$[(\alpha^{i_1}, \alpha^{j_1})] \cup [(\alpha^{i_2}, \alpha^{j_2})] \cup \cdots \cup [(\alpha^{i_s}, \alpha^{j_s})]$,
		\end{center}
		where $\alpha$ is a primitive element of $\mathbb{F}_{q^m}$ and $i_\gamma, j_\gamma$ are elements of the set $\{1, 2, \dots, q^m - 2\}$ for all $\gamma = 1, 2, \dots, s$. Assume that the pairs $(\alpha^{i_\gamma}, \alpha^{j_\gamma})$ are representatives of distinct $\mathbb{F}_q$-conjugacy
		classes. Let $D$ be the $2$-D cyclic code of the same area over $\mathbb{F}_{q^m}$ whose dual’s zero set is
		\begin{center}
			$\{(\alpha^{i_1}, \alpha^{j_1}), (\alpha^{i_2}, \alpha^{j_2}), \dots, (\alpha^{i_s}, \alpha^{j_s})\}$.
		\end{center}
		Then we have the following representations for the codes $D$ and $C$; where $\lambda_\gamma$ runs
		through $\mathbb{F}_{q^m}$ for every $\gamma = 1, 2, \dots, s$:
		\begin{align*}
			D & =\left\{\sum_{\gamma = 1}^{s} \lambda_\gamma v_\gamma\right\}\\
			& = \left\{\begin{pmatrix}
					\lambda_1x^{j_1} + \cdots + \lambda_sx^{j_s}\\
					\lambda_1 \alpha^{i_1}x^{j_1} + \cdots + \lambda_s \alpha^{i_s}x^{j_s}\\
					\vdots\\
					\lambda_1 (\alpha^{i_1})^{q^m - 2}x^{j_1} + \cdots + \lambda_s (\alpha^{i_s})^{q^m - 2}x^{j_s}
			\end{pmatrix}_{x \in A}\right\}\\
		& = \left\{(\lambda_1 (\alpha^{i_1})^\delta x^{j_1} + \cdots + \lambda_s (\alpha^{i_s})^\delta x^{j_s})_{x \in A, \delta \in \mathcal{I}}\right\}\\
		& = \left\{(\lambda_1 x^{i_1} + \cdots + \lambda_s x^{i_s}, \lambda_1 \alpha^{j_1}x^{i_1} + \cdots + \lambda_s \alpha^{j_s}x^{i_s}, \cdots )_{x \in A}\right\}\\
		& = \left\{(\lambda_1 (\alpha^{j_1})^\delta x^{i_1} + \lambda_2 (\alpha^{j_2})^\delta x^{i_2} + \cdots + \lambda_s (\alpha^{j_s})^\delta x^{i_s})_{x \in A, \delta \in \mathcal{I}}\right\},
	\end{align*}
	
	\begin{align*}
		C & =\left\{\sum_{\gamma = 1}^{s} \text{\bf tr}(\lambda_\gamma v_\gamma)\right\}\\
		& = \left\{\begin{pmatrix}
			\mathrm{tr}(\lambda_1x^{j_1} + \cdots + \lambda_sx^{j_s})\\
			\mathrm{tr}(\lambda_1 \alpha^{i_1}x^{j_1} + \cdots + \lambda_s \alpha^{i_s}x^{j_s})\\
			\vdots\\
			\mathrm{tr}(\lambda_1 (\alpha^{i_1})^{q^m - 2}x^{j_1} + \cdots + \lambda_s (\alpha^{i_s})^{q^m - 2}x^{j_s})
		\end{pmatrix}_{x \in A}\right\}\\
		& = \left\{(\mathrm{tr}(\lambda_1 (\alpha^{i_1})^\delta x^{j_1} + \cdots + \lambda_s (\alpha^{i_s})^\delta x^{j_s}))_{x \in A, \delta \in \mathcal{I}}\right\}\\
		& = \left\{(\mathrm{tr}(\lambda_1 x^{i_1} + \cdots + \lambda_s x^{i_s}), \mathrm{tr}(\lambda_1 \alpha^{j_1}x^{i_1} + \cdots + \lambda_s \alpha^{j_s}x^{i_s}), \cdots )_{x \in A}\right\}\\
		& = \left\{(\mathrm{tr}(\lambda_1 (\alpha^{j_1})^\delta x^{i_1} + \lambda_2 (\alpha^{j_2})^\delta x^{i_2} + \cdots + \lambda_s (\alpha^{j_s})^\delta x^{i_s}))_{x \in A, \delta \in \mathcal{I}}\right\}.
	\end{align*}
	\end{thm}

	\begin{coro}\textup{(See Güneri \cite{guneri2004artin})}
		Consider the code $C$ of area $(q^m - 1) \times (q^m - 1)$ over $\mathbb{F}_q$ whose dual has as a basic zero set
		\begin{center}
			$\mathrm{BZ}(C^\perp) = \{(\alpha^{i_1}, \alpha^{j_1}), (\alpha^{i_2}, \alpha^{j_2}), \dots, (\alpha^{i_s}, \alpha^{j_s})\}$.
		\end{center}
		Let $\tilde{C}$ be the code of same area over $\mathbb{F}_q$ for which the dual has as a basic zero set
		\begin{center}
			$\mathrm{BZ}(\tilde{C}^\perp) = \{(\alpha^{j_1}, \alpha^{i_1}), (\alpha^{j_2}, \alpha^{i_2}), \dots, (\alpha^{j_s}, \alpha^{i_s})\}$.
		\end{center}
		Then the weight enumerators of $C$ and $\tilde{C}$ are the same.
	\end{coro}

	\paragraph*{} The whole weight enumerator of $C$ is related to the following family:
	\begin{center}
		$\mathcal{F} = \{y^q - y = \lambda_1x^{j_1} + \lambda_2x^{j_2} + \cdots + \lambda_sx^{j_s}; \, \lambda_j \in \mathbb{F}_{q^m}\}$.
	\end{center}
	
	\begin{thm}\textup{(See Güneri \cite{guneri2004artin})}
		Let $C$ be the $2$-D cyclic code of area $(q^m - 1) \times (q^m - 1)$ over $\mathbb{F}_q$ whose dual has as a basic zero set
		\begin{center}
			$\mathrm{BZ}(C^\perp) = \{(\alpha^{i_1}, \alpha^{j_1}), (\alpha^{i_2}, \alpha^{j_2}), \dots, (\alpha^{i_s}, \alpha^{j_s})\} \subset \Omega$.
		\end{center}
		Assume that the $j_\gamma$'s are distinct and the $q$-cyclotomic coset mod $q^m - 1$ containing each
		$j_\gamma$ has cardinality $m = [\mathbb{F}_{q^m} : \mathbb{F}_q]$. Then
		\begin{enumerate}
			\item $\dim_{\mathbb{F}_q}(C) = sm$.
			\item If $d$ denotes the minimum distance of $C$, we have
			\begin{center}
				$d \geq (q^m - 1)\left(q^m - \frac{N}{q}\right)$,
			\end{center}
		\end{enumerate}
		where $N$ is in the set $\{q, 2q, \dots, (q^m - 1)q\}$ and it is the maximum number of affine $\mathbb{F}_{q^m}$-rational points that a nontrivial member in the family $\mathcal{F} = \{y^q - y = \lambda_1x^{j_1} + \lambda_2x^{j_2} + \cdots + \lambda_sx^{j_s}; \, \lambda_\gamma \in \mathbb{F}_{q^m}\}$ can have.
	\end{thm}

	\begin{coro}\textup{(See Güneri \cite{guneri2004artin})}
		Let $C$ be the $2$-D cyclic code of area $(q^m - 1) \times (q^m - 1)$ over $\mathbb{F}_q$ whose dual has as a basic zero set
		\begin{center}
			$\mathrm{BZ}(C^\perp) = \{(\alpha^{i_1}, \alpha^{j_1}), (\alpha^{i_2}, \alpha^{j_2}), \dots, (\alpha^{i_s}, \alpha^{j_s})\}$.
		\end{center}
		Assume that the $q$-cyclotomic coset containing each
		$j_\gamma$ has cardinality $m = [\mathbb{F}_{q^m} : \mathbb{F}_q]$.
		For every $\gamma = 1, 2, \dots s$, write $j_\gamma$ as $j_\gamma = r_\gamma p^{\eta_\gamma}$, where $p$ does not divide $r_\gamma$. Suppose the
		$r_\gamma$'s are all distinct and let $r = \max\{r_1, r_2, \dots, r_s\}$. Then $\dim_{\mathbb{F}_q}(C) = sm$ and if $d$ denotes
		the minimum distance of $C$; we have
		\begin{center}
			$d \geq (q^m - 1)\left(q^m - \frac{N}{q}\right)$,
		\end{center}
		where $N$ is the maximum of the set $\{q, 2q, \dots, (q^m - 1)q\}$ that is less than or equal to
		\begin{center}
			$q^m + \frac{(q - 1)(r - 1)}{2}\left[2\sqrt{q^m}\right]$.
		\end{center}
	\end{coro}
	
	\begin{thm}\textup{(See Güneri \cite{guneri2004artin})}
		Let $C$ be the code over $\mathbb{F}_q$ of area $(q^m - 1) \times (q^m - 1)$ whose dual has as a basic zero set
		\begin{center}
			$\mathrm{BZ}(C^\perp) = \{(\alpha^{i_1}, \alpha), (\alpha^{i_2}, \alpha)\}$.
		\end{center}
	Then $C$ is of dimension $2m$ over $\mathbb{F}_q$ and if $\theta$ denotes the order of $\alpha^{i_2 - i_1}$ in the multiplicative group $\mathbb{F}_{q^m}^\ast$; then the weights and their frequencies for $C$ are
	\begin{center}
		\begin{tabular}{cc}
			weight & frequency \medskip\\
			$\left(q^m - 1 - \frac{q^m - 1}{\theta}\right)(q^m - q^{m - 1})$ & $\theta \cdot (q^m - 1)$ \medskip\\
			$(q^m - 1)(q^m - q^{m - 1})$ & $q^{2m} - \theta \cdot (q^m - 1) - 1$.
		\end{tabular}
	\end{center}
	\end{thm}

	\begin{prop}\textup{(See Güneri \cite{guneri2004artin})}
		Let $C$ be the code over $\mathbb{F}_q$ of area $(q^m - 1) \times (q^m - 1)$ whose dual has as a basic zero set
		\begin{center}
			$\mathrm{BZ}(C^\perp) = \{(\alpha^{i_1}, \alpha^j), (\alpha^{i_2}, \alpha^j)\}$.
		\end{center}
		Let $i \neq j = rp^\eta$; where $p$ does not divide $r$; and assume that the $q$-cyclotomic coset
		containing $j \mod q^m - 1$ has cardinality $m$. Then
		\begin{center}
			$d \geq \left(q^m - 1 - \frac{q^m - 1}{\theta}\right)\left(q^m - \frac{N}{q}\right)$,
		\end{center}
		where $d$ is the minimum distance of $C$, $N$ is the maximum of $\{q, 2q, \dots, (q^m - 1)q\}$
		which is less than or equal to
		\begin{center}
			$q^m + \frac{(q - 1)(r - 1)}{2}\left[2\sqrt{q^m}\right]$,
		\end{center}
	and $\theta$ is the order of $\alpha^{i_2 - i_1}$ in the multiplicative group $\mathbb{F}_{q^m}^\ast$.
	\end{prop}

	\begin{prop}\textup{(See Güneri \cite{guneri2004artin})}
		Let $C$ be the code over $\mathbb{F}_q$ of area $(q^m - 1) \times (q^m - 1)$ whose dual has as a basic zero set
		\begin{center}
			$\mathrm{BZ}(C^\perp) = \{(\alpha^{i_1}, \alpha^{j_1}), (\alpha^{i_2}, \alpha^{j_2}), (\alpha^{i_3}, \alpha^{j_2})\}$.
		\end{center}
		Let $j_\gamma = r_\gamma p^{\eta_\gamma} \, (\gamma = 1, 2)$, where $p$ does not divide $r_\gamma$; and suppose that the $q$-cyclotomic coset
		containing $j_\gamma \mod q^m - 1$ has cardinality $m$. Let $r = \max{r_1, r_2}$. If $r_1$ and $r_2$ are distinct, then
		\begin{center}
			$d \geq \left(q^m - 1 - \frac{q^m - 1}{\theta}\right)\left(q^m - \frac{N}{q}\right)$,
		\end{center}
		where $d$ is the minimum distance of $C$, $N$ is the maximum of $\{q, 2q, \dots, (q^m - 1)q\}$
		which is less than or equal to
		\begin{center}
			$q^m + \frac{(q - 1)(r - 1)}{2}\left[2\sqrt{q^m}\right]$,
		\end{center}
		and $\theta$ is the order of $\alpha^{i_3 - i_2}$ in the multiplicative group $\mathbb{F}_{q^m}^\ast$.
	\end{prop}

		\begin{prop}\textup{(See Güneri \cite{guneri2004artin})}
		Let $C$ be the code over $\mathbb{F}_q$ of area $(q^m - 1) \times (q^m - 1)$ whose dual has as a basic zero set
		\begin{center}
			$\mathrm{BZ}(C^\perp) = \{(\alpha^{i_1}, \alpha^{j_1}), (\alpha^{i_2}, \alpha^{j_1}), (\alpha^{i_3}, \alpha^{j_2}), (\alpha^{i_4}, \alpha^{j_2})\}$.
		\end{center}
		Let $j_\gamma = r_\gamma p^{\eta_\gamma} \, (\gamma = 1, 2)$, where $p$ does not divide $r_\gamma$; and suppose that the $q$-cyclotomic coset
		containing $j_\gamma \mod q^m - 1$ has cardinality $m$. Let $r = \max\{r_1, r_2\}$, $\theta$ be the order of $\alpha^{i_2 - i_1}$ in $\mathbb{F}_{q^m}^\ast$, and assume this is the same as the order of $\alpha^{i_4 - i_3}$  If $r_1$ and $r_2$ are distinct, then
		\begin{center}
			$d \geq \left(q^m - 1 - \frac{q^m - 1}{\theta}\right)\left(q^m - \frac{N}{q}\right)$,
		\end{center}
		where $d$ is the minimum distance of $C$, $N$ is the maximum of $\{q, 2q, \dots, (q^m - 1)q\}$
		which is less than or equal to
		\begin{center}
			$q^m + \frac{(q - 1)(r - 1)}{2}\left[2\sqrt{q^m}\right]$.
		\end{center}
	\end{prop}

	\section{Quasi-cyclic product codes}

	\paragraph*{} The papers \cite{zeh2015construction, zeh2016spectral} by Zeh and Ling considered a linear quasi-cyclic product code of two given quasi-cyclic codes of relatively prime lengths over finite fields. This is a generalization of work of Burton and Weldon \cite{burton1965cyclic} for quasi-cyclic product codes.
	
	\paragraph*{} We can represent a codeword of an $[l.m,k,d]_{q}$ $l$-quasi-cyclic code $C$ as\\
	$c(X)=(c_{0}(X), \dots, c_{l-1}(X)) \in \mathbb{F}_{q}[X]^{l}$ where each entry is given by 
	\begin{equation}
	    c_{j}(X)=\sum_{i=0}^{m-1} c_{j,i} X^i,~ \forall j \in [l):=[0,l).
	\end{equation}
	
	\begin{lemma}\textup{(See Zeh and Ling \cite{zeh2016spectral})(Codeword Representation: Vector to Univariate polynomial)}
	    Let $c(X)=(c_{0}(X), \dots, c_{l-1}(X))$ be a codeword of an $[l.m,k,d]_{q}$ $l$-quasi-cyclic code $C$, where the entries are defined as in $(3)$ below. Then a codeword in $C$, represented as one univariate polynomial of degree smaller than $lm$, is 
	    \begin{equation}
	        c(X)=\sum_{j=0}^{l-1} c_{j}(X^l) X^j.
	    \end{equation}
	\end{lemma}
	Lally and Fitzpatrick showed that an $l$-quasi-cyclic code $C$ can be represented as an $R$-submodule of algebra $R^l$, where $R=\mathbb{F}_{q}[X]/ \langle X^m - 1 \rangle$. The code $C$ is the image of an $\mathbb{F}_{q} [X]$-submodule $\tilde{C}$ of $\mathbb{F}_{q} [X]^{l}$ containing $\tilde{K} = \langle(X^m - 1)e_ j , j \in [l) \rangle$, where $e_j \in  \mathbb{F}_{q} [X]^l$ is the standard basis vector with one in position $j$ and zero elsewhere under the natural homomorphism
	\begin{equation}
    	\phi: \mathbb{F}_{q}[X]^l \rightarrow R^l ~(c_{0}(X), \dots, c_{l-1}(X)) \mapsto (c_{0}(X)+ \langle X^m-1 \rangle, \dots, c_{l-1}(X)+\langle X^m-1 \rangle).
	\end{equation}
	The submodule has a generating set of the form $\{u_i,~ i \in [z),~ (X^m - 1)e_j , j \in [l)\}$, where $u_i \in  \mathbb{F}_{q} [X]^l$ and $z \leq l$ and can be represented as a matrix with entries in $\mathbb{F}_{q} [X]$:
	\[
	U(X)=\begin{bmatrix}
		u_{0,0}(X) &u_{0,1}(X) &\dots &u_{0,l-1}(X)\\
		u_{1,0}(X) &u_{1,1}(X) &\dots &u_{1,l-1}(X)\\
		\vdots &\vdots &\ddots &\vdots\\
		u_{z-1,0}(X) &u_{z-1,1}(X) &\dots &u_{z-1,l-1}(X)\\
		X^m - 1 &~ &~&~\\
		~ &X^m -1 &~ &0\\
		~&0 &\ddots &~\\
		~&~&~& X^m-1
	\end{bmatrix}
	\]
	Every matrix $U(X)$ of an $l$-quasi-cyclic code $C$ can be transformed to a reduced Groebner basis (RGB) with respect to the position-over-term order (POT) in $\mathbb{F}_q [X]^l$. A basis in RGB/POT form can be represented by an upper-triangular $l \times l$ matrix with entries in $\mathbb{F}_q [X]$ as follows:
	\begin{equation}
		G(X)=\begin{bmatrix}
			g_{0,0}(X) &g_{0,1}(X) &\dots &g_{0,l-1}(X)\\
			~ &g_{1,1}(X) &\dots &g_{1,l-1}(X)\\
			~ &0 &\ddots &\vdots\\
			~ &~ &~&g_{l-1,l-1}(X)
	\end{bmatrix}
	\end{equation}
	where the following conditions must be fulfilled:\\
	C1 : $g_{i, j} (X)=0$, $\forall ~0 \leq j <l$,\\
	C2 : $deg ~g_{ j,i} (X)< deg ~g_{i,i} (X)$, $\forall ~ j<i$, $i\in [l)$,\\
	C3 : $g_{i,i} (X)|(X^m-1)$, $\forall ~i \in [l)$,\\
	C4 : if $g_{i,i} (X)=X^m-1$ then $g_{i, j} (X)=0$, $\forall ~j \in [i + 1, l)$.\\
	We refer to these conditions as RGB/POT conditions C1-C4.\\

	Assume $gcd(m,$ char $\mathbb{F}_{q})=1$.

	\paragraph*{} We call an $[lm, k, d]_q$-quasi-cyclic code $C$ an $r$-level quasi-cyclic code if there is an index $r \in [l)$ for which the RGB/POT matrix as defined above is such that $g_{r-1,r-1}(X)\neq X^m-1$ and $g_{r,r} (X) =\dots = g_{l-1,l-1}(X) = X^m-1$.
	
	\paragraph*{} The generator matrix in RGT/POT form of an $[ lm, k, d]_q$ $1$-level quasi-cyclic code is given by the following result.
	
	\begin{prop}\textup{(See Zeh and Ling \cite{zeh2016spectral})}
    The generator matrix in RGB/POT form of an $[lm, k, d]_q$ $1$-level $l$-quasi-cyclic code $C$ has the following form:
	$$G(X) = (g(X)~ g(X) f_{1}(X) \dots g(X) f_{l-1}(X)),$$
	where $g(X)|(X^m - 1)$, $deg~ g(X) = m - k$, and $f_1(X), \dots, f_{l - 1}(X) \in \mathbb{F}_{q} [X]$.
	\end{prop}

	\subsection{Spectral analysis of quasi-cyclic codes}
	Let $G(X)$ be the upper-triangular generator matrix of a given $[lm, k, d]_q$ $l$-quasi-cyclic code C in RGB/POT form. Let $\alpha \in \mathbb{F}_{q^s}$ be an $m$th root of unity. An eigenvalue $\lambda_i$ of $C$ is defined to be a root of $det(G(X))$, i.e\., a root of $\prod_{j=0}^{l-1} g_{j,j}(X)$. The algebraic multiplicity of $\lambda_i$ is the largest integer $u_i$ such that $(X-\lambda_i )^{u_i} | det(G(X))$. The geometric multiplicity of an eigenvalue $\lambda_i$ is defined as the dimension of the right kernel of the matrix $G(\lambda_i )$. The solution space is called the right kernel eigenspace and it is denoted by $\mathcal{V}_{i}$. Furthermore, it was shown that, for a matrix $G(X)\in \mathbb{F}_{q} [X]^{l \times l}$ in RGB/POT form, the algebraic multiplicity $u_i$ of an eigenvalue $\lambda_{i}$ equals the geometric multiplicity
	
	\paragraph*{} A generator matrix $G(X)$ of $C$ that satisfies RGB/POT conditions C1, C3 and C4, but not C2, is called a matrix in Pre-RGB/POT form.
	More explicitly, the generator matrix has the following form:
	\[
	\bar{G}(X)=\begin{bmatrix}
	g_{0,0}(X) &\bar{g}_{0,1}(X) &\dots &\bar{g}_{0,l-1}(X)\\
	~ &g_{1,1}(X) &\dots &\bar{g}_{1,l-1}(X)\\
	~ &0 &\ddots &\vdots\\
	~ &~ &~&g_{l-1,l-1}(X)
	\end{bmatrix}
	\]
	where the entries of $\bar{G}(X)$ that can be different from their
	counterparts in the RGB/POT form, are marked by a bar.
	
	\begin{lemma}\textup{(See Zeh and Ling \cite{zeh2016spectral})(Equivalence of the Spectral Analysis of a Matrix
	in Pre-RGB/POT Form)}
	 Let $G(X)$ be an $l \times l$ generator matrix of an $l$-quasi-cyclic code $C$ in RGB/POT form and let $\bar{G}(X)$ be a generator matrix of the same code in Pre-RGB/POT form. Let $\lambda_i$ be an eigenvalue of $G(X)$. Then, the right kernels of
	$G(\lambda_i)$ and $\bar{G}( \lambda_i)$ are equal, i.e.\, the (algebraic and geometric) multiplicity and the corresponding eigenvalues are identical.   
	\end{lemma}

	\paragraph*{} Let $\mathcal{V} \subseteq \mathbb{F}_{q^s}^l$ be an eigenspace. Define the $[n^{ec} = l, k^{ec}, d^{ec}]_q$ eigencode corresponding to $\mathcal{V}$ by
	$$C(\mathcal{V}):=\{c \in \mathbb{F}_{q}^{l}: \forall v \in \mathcal{V}, v \circ c=0\}.$$
	
	\subsection{Quasi-cyclic product codes}
	Let $A$ be an $[n_A =l_A m_A, k_A, d_A]_q$ $l_{A}$-quasi-cyclic code generated by the following matrix in RGB/POT form as defined in:
	\[
	G_{A}(X)=\begin{bmatrix}
	g_{0,0}^{A}(X) &{g}^{A}_{0,1}(X) &\dots &{g}^{A}_{0,l-1}(X)\\
	~ &g^{A}_{1,1}(X) &\dots &{g}^{A}_{1,l-1}(X)\\
	~ &0 &\ddots &\vdots\\
	~ &~ &~&g^{A}_{l-1,l-1}(X)
	\end{bmatrix}
	\]
	and let $B$ be an $[n_B = l_B m_B, k_B, d_B]_q$ $l_{B}$-quasi-cyclic code with generator matrix in RGB/POT form:
	\[
	G_{B}(X)=\begin{bmatrix}
	g_{0,0}^{B}(X) &{g}^{B}_{0,1}(X) &\dots &{g}^{B}_{0,l-1}(X)\\
	~ &g^{B}_{1,1}(X) &\dots &{g}^{B}_{1,l-1}(X)\\
	~ &0 &\ddots &\vdots\\
	~ &~ &~&g^{B}_{l-1,l-1}(X)
	\end{bmatrix}
	\]
	 \begin{lemma}\textup{(See Zeh and Ling \cite{zeh2016spectral})(Mapping to a Univariate Polynomial)}: Let
	$A$ be an $l_{A}$-quasi-cyclic code of length $n_{A}$ and let $B$ be an $l_{B}$-quasi-cyclic code of length $n_B$. The product code $A \otimes B$ is an $l_A l_B$-quasi-cyclic code of length $n = n_A n_B$ if $gcd(n_A, n_B) = 1$.
	\end{lemma}
	
	Instead of representing a codeword in $A\otimes B$ as one univariate polynomial in $\mathbb{F}_{q} [X]$, we want to represent it as a vector of $l_{A} l_{B}$ univariate polynomials in $\mathbb{F}_{q} [X]$ to obtain an explicit expression of the basis
	of the $l_A l_B$-quasi-cyclic product code $A \otimes B$.
	
	\begin{lemma}\textup{(See Zeh and Ling \cite{zeh2016spectral})(Mapping to $l_A l_B$ Univariate Polynomials)}
	 Let $A$ be an $l_A$-quasi-cyclic code of length $n_A = l_A m_A$ and let $B$ be an $l_B$-quasi-cyclic code of length $n_B = l_B m_B$. Let $l = l_A l_B$, $m = m_A m_B$, and $n = n_A n_B$. Let $(m_{i, j} )^{j\in [n_A)}_{i\in [n_B)}$ be a codeword of the $l$-quasi-cyclic product code $A \otimes B$ where each row is in $A$ and each column is in $B$. Define $l$ univariate polynomials as: $c_{g,h}(X) \equiv X^{\nu(g,h)} \sum_{i=0}^{m_{B}-1} \sum_{j=0}^{m_A-1} m_{i l_B+g, j l_A+h} X^{\bar{\mu}(i, j )} \mod (X^m -1)$, $\forall g \in [l_B)$, $h \in  [l_A), $
	with $\nu(g, h) = g(-b m_B) + h(-a m_A) \mod m$, $\bar{\mu}(i, j) = ia n_A + jb n_B \mod m.$ Then the codeword $c(X) \in  A \otimes B$ corresponding to $(m_{i, j})^{j\in [n_A)}_{i \in [n_B)}$ is given by: $c(X) \equiv \sum_{g=0}^{l_B-1}\sum_{h=0}^{l_A -1} c_{g,h}(X^{l_A l_B} )X^{g l_A+h l_B} \mod (X^n-1). $  
	    
	\end{lemma}
	
	The following theorem gives the generator matrix in
	RGB/POT form of an $l$-quasi-cyclic product code $A \otimes B$ where the row-code $A$ is a 1-level $l$-quasi-cyclic code and $B$ is a cyclic code.
	
	\begin{thm}\textup{(See Zeh and Ling \cite{zeh2016spectral})(Generator Matrix of a $1$-level Quasi-Cyclic Product Code in RGB/POT Form)}
	Let $A$ be an $[n_A = l_A m_A, k_A, d_A]_q$ $1$-level $l$-quasi-cyclic code with generator matrix in RGB/POT form:
	$$G^{A}(X) = (g^{A}_{0,0}(X) g^{A}_{0,1}(X) \dots g^{A}_{0,l-1}(X))
	= (g_A(X) g_A(X) f^{A}_{1} (X) \dots g^{A}(X) f^{A}_{l-1}(X)).$$ Let $B$ be an $[n_B = m_B, k_B, d_B]_q$
	cyclic code with generator polynomial $g_B(X) \in \mathbb{F}_{q} [X]$. Let $m = m_{A} m_{B}$. Then the generator matrix of the $1$-level $l$-quasi-cyclic product code $A \otimes B$ in RGB/POT form is:
	$$G(X) = (g(X) g(X) f^{A}_{1}(X^{b n_B} ) \dots g(X) f^A_{l-1}(X^{b n_B} ))\cdot \text{diag} \, (1, X^{-a m_A}, X^{-2 a m_A},\dots, X^{-(l-1)a m_A}),$$
	where $g(X) = gcd (X^m -1, g^A(X^{b n_B} )g^B(X^{a n_A}))$.
	\end{thm}
	The authors also conjectured the (general form of the) generator matrix in Pre-RGB/POT form of an $l_A l_B$-quasi-cyclic product code $A \otimes B$.
	
	\subsection{Spectral analysis of quasi-cyclic product codes}
	Let $A$ be an $[n_A = l m_{A}, k_A, d_A]_q$ $l$-quasi-cyclic code with generator matrix in RGB/POT form and let $B$ be an $[n_B = l_B m_B, k_B, d_B]_q$ cyclic code with generator polynomial $g_{B}(X)$. Let $m = m_{A}m_{B}$. The product code $A \otimes B$ is an $[lm, k_{A}k_{B}, d_{A}d_{B}]_q$ $l$-quasi-cyclic code with generator matrix in Pre-RGB/POT form i.e., their entries are:
	$$g_{h,h}(X) = u_{h}(X)(X^m-1)+v_{h} (X)g^{A}_{h,h}(X^{b n_B})g^{B}(X^{a n_A} ),~\forall h \in  [l),$$
	$$\bar{g}_{h,h'} (X) = v_{h} (X)g^{A}_{h,h'} (X^{b n_B} )g^{B}(X^{a n_A} ),~ \forall h \in  [l), h' \in [h + 1, l).$$
	Furthermore, let throughout two nonzero integers $a, b$ be such that $a n_A + b n_B = 1$. For a given set $A = \{a_ 0, a_1,\dots, a_{|A|-1}\}$, denote by $A^{\otimes z}=\{a_i +z | a_i \in A\}.$
	\begin{lemma}\textup{(See Zeh and Ling \cite{zeh2016spectral})(Eigenvalues Of Maximal Multiplicity)}
	Let $A$ be an $[l m_A, k_A, d_A]_q$ $l$-quasi-cyclic code with generator matrix $G^{A}(X)$ in RGB/POT form. Let $\alpha$ be an element of order $m_A$ in $\mathbb{F}_{q^{s_A}}$, $B$ an $[n_B = m_B, k_B, d_B]_q$ cyclic code, and $\beta$ an element of order $m_B$ in $\mathbb{F}_{q^{s_B}} $. Define $s= lcm(sA, sB)$.
	Let $\gamma=\alpha \beta$ be in $\mathbb{F}_{q^s}$. Let the set $A^{(l)} \subseteq [m_A)$ contain the exponents of all eigenvalues $\lambda^{A}_{z}= \alpha^z$,~ $\forall z \in A^{(l)}$ of $A$ of
	(algebraic and geometric) multiplicity $l$. Let $B \subseteq [m_B)$ be the defining set of $B$, i.e\., the set of exponents of all roots of the generator polynomial $g_{B}(X) = \prod_{i\in B}(X-\beta_{i})$ of $B$. Then, the set:
	\begin{align*}
	C^{(l)} &= A^{(l)} \cup  A^{{(l)}^{\otimes m_A}} \cup A^{{(l)}^{\otimes 2 m_A}} \cup \dots\\
	& \cup A^{{(l)}^{\otimes (m_B-1) m_A}} \cup B \cup B^{\otimes m_{B}} \cup \dots\\
	&\cup B^{\otimes (m_A-1)m_B}
	\end{align*}
	is the set of all the exponents of the eigenvalues $\lambda_z= \gamma^z$ for all $z \in C^{(l)}$ of the $l$-quasi-cyclic product code $A \otimes B$ of maximal multiplicity $l$. Furthermore, we have $|C^{(l)}| = |A^{(l)}|m_B + (m_B-k_B)m_A-|A^{(l)}|(m_B-k_B) = (m_B-k_B)m_A +|A^{(l)}|k_B$.
	\end{lemma}
	
	The following lemma considers eigenvalues of the $l$-quasi-cyclic
	product code $A \otimes B$ of multiplicity smaller than $l$
	and their corresponding eigenvectors. 
	\begin{lemma}\textup{(See Zeh and Ling \cite{zeh2016spectral})(Eigenvalues of Smaller Multiplicity and Their
	Eigenvectors)} Let the two codes $A$ and $B$ with parameters be
	given as in previous lemma. Let the set $A^{(r)} \subseteq [m_A)$ contain the exponents of all eigenvalues $\lambda^{A}_{z}=\alpha z$, $\forall~ z \in  A^{(r)}$ of $A$ of (algebraic and
	geometric) multiplicity $r \in [l)$. Let $v^{A}_{z,0}, v^{A}_{z,1}, \dots , v^{A}_{z,r-1} \in  \mathbb{F}_{q^{s_A}}$ be the corresponding $r$ eigenvectors of $\lambda^{A}_{z}$, i.e.\, a basis of the right kernel of $G^{A}(\lambda^{A}_{z})$. Let $B \subseteq [m_B)$
	be the defining set of $B$, i.e.\, the set of exponents of all roots of the generator polynomial $g_{B}(X)=\prod_{i\in B}(X-\beta_{i} )$ of $B$. Let $\gamma=\alpha \beta$ be in $\mathbb{F}_{q^s}$. Then, the set:
	\begin{align*}
	C^{(r)} = &(A^{(r)} \cup A^{(r)^{\otimes m_A}} \cup A^{(r)^{2 \otimes m_A}} \cup \dots \cup A^{(r)^{\otimes(m_{B}-1) m_A}})\\
	& \backslash (B \cup B^{\otimes m_B} \cup B^{2 \otimes m_B} \cup \dots \cup B^{\otimes(m_{A}-1) m_{B}})
	\end{align*}
	is the set of all exponents of the eigenvalues $\lambda_{z}=\gamma^{z}$ for all $z \in C^{(r)}$ of the $l$-quasi-cyclic product code $A \otimes B$ of multiplicity $r$. The number of eigenvalues of $A \otimes B $ of multiplicity
	$r$ is $|C^{(r)}| = |A^{(r)}| k_B$. Furthermore, the corresponding
	eigenvectors $v_{z,0}, v_{z,1},\dots , v_{z,r-1}$ are:
	$$v_{z, j} = v^{A}_{z \mod m_{A, j}}, ~\forall z \in C^{(r)}, ~j \in [r).$$
	\end{lemma}
	
	The following theorem gives a new lower bound on the minimum Hamming distance of a given $l$-quasi-cyclic code $A$ by embedding
	it into an $l$-quasi-cyclic product code $A$.
	\begin{thm}\textup{(See Zeh and Ling \cite{zeh2016spectral})(Generalized Semenov–Trifonov Bound)}
	Let $A$ be an $[l m_A, k_A, d_A]_q$ $l$-quasi-cyclic code, $\alpha$ an element of order $m_A$ in $\mathbb{F}_{q^{s_A}}$ , $B$ an $[n_B = m_B, k_B, d_B]_q$ cyclic code, and $\beta$ an element of order $m_B$ in $\mathbb{F}_{q^{s_B}}$ . Furthermore, let $gcd(m_A,m_B) = 1$. Let the integers $f_1 \geq  0$, $f_2 \geq 0$, $z_1 > 0$, $z_2 > 0$, $\delta>2$ with $gcd(z_1,m_A) = 1$, and $gcd(z_2,m_B) = 1$ be given, such that:
	$$\sum_{i=0}^{\infty} (a(\alpha^{ f_1+i z_1} ) b(\beta^{ f_2+i z_2} )\, \circ \, v X^i \equiv 0 \mod X^{\delta-1}$$ holds for all $a(X) = (a_{0}(X) ~a_{1}(X) \dots a_{l-1}(X)) \in A$, $b(X) \in B$, and for all $v = (v_0 ~ v_1 ~\dots~ v_{l-1})\in \mathbb{F}_{q^{s_A}}$ in the intersection of the eigenspaces
	$$\mathcal{V}= \cap_{j \in D} \mathcal{V}_{j}$$
	where
	$$D=\{ f_{1} + i z_{1} | b(\beta^{ f_2+i z_2} ) \neq 0,~ \forall ~i \in  [\delta - 1)\}. $$
	Let the distance of the eigencode $\mathbb{C}(\mathcal{V})$ be $d^{ec}$. Then:
	$$ d_{A} \geq d^{*}:=\Bigg\lceil \frac{min(\delta, d^{ec})}{d_{B}} \Bigg\rceil$$
	\end{thm}

	\section{2-D skew-cyclic codes over $\mathbb{F}_q [x, y; \rho, \theta]$}
	
	\paragraph*{} The theory of two-dimensional cyclic codes was generalized to 2-D skew-cyclic codes in \cite{li20142-D} by X. Li and H. Li considering the relationship between left $\mathbb{F}_q[x, y; \rho, \theta]$-submodules of the left $\mathbb{F}_q[x, y; \rho, \theta]$-module $\mathbb{F}_q[x, y; \rho, \theta]/\langle x^s-1, y^l-1\rangle_l$ and 2-D skew-cyclic codes of size $s \times l$, where $\langle x^s-1, y^l-1\rangle_l$ is the left ideal generated by $x^s-1$ and $y^l-1$ in $\mathbb{F}_q[x, y; \rho, \theta]$.
	
	\paragraph*{} Consider the finite field $\mathbb{F}_q$ and two of its automorphisms $\rho$ and $\theta$. Then we have a set of formal bivariate skew polynomials
	\begin{center}
		$\mathbb{F}_q[x, y; \rho, \theta]= \{\sum \sum a_{ij}x^iy^j \, | \, a_{ij} \in \mathbb{F}_q, (i, j) \in \mathbb{Z}^2_+\}.$
	\end{center}
	
	\paragraph*{} The addition in $\mathbb{F}_q[x, y; \rho, \theta]$ is the usual polynomial addition and the multiplication is defined as
	\begin{center}
		$ax^iy^j \ast bx^ry^t = a \rho^i\theta^j(b)x^{i+r}y^{j+t}.$
	\end{center}
	These rules are extended to all elements of $\mathbb{F}_q[x, y; \rho, \theta]$ by associativity and distributivity showing that $\mathbb{F}_q[x, y; \rho, \theta]$ is a ring.
	
	\paragraph*{} Let $C$ be a linear code over $\mathbb{F}_q$ of length $sl$ whose codewords are viewed as $s \times l$ arrays, i.e. $c \in C$ is written as
			$$c =	
			\begin{pmatrix}
				c_{0,0} & \cdots & c_{0,l-1}\\
				c_{1,0} & \cdots & c_{1,l-1}\\
				& \vdots &			\\
				c_{s-1,0} & \cdots & c_{s-1,l-1}
			\end{pmatrix}.$$
		
		\paragraph*{} If $C$ is closed under row $\rho$-skew-shift and column $\theta$-skew-shift of codewords, then we call $C$ a 2-D skew-cyclic code of size $s \times l$ over $\mathbb{F}_q$ under $\rho$ and $\theta$.
	
	\paragraph*{} Clearly, a 2-D skew-cyclic code is a 2-D cyclic code when $\rho$ and $\theta$ are all identity mappings. Now a description of the bivariate skew polynomial ring $\mathbb{F}_q[x, y; \rho, \theta]$ is provided.
	
	\paragraph*{} Let $\mathbb{Z}^2_+ = \{(i, j) \, | \, i, j \geq 0, i, j \in \mathbb{Z}\}$. $\mathbb{Z}^2_+$ is a partial ordered set with $(i, j) \geq (k,l)$ if and only if $i \geq k$ and $j \geq l$.
	
	\paragraph*{} $\mathbb{Z}^2_+$ can be also totally ordered by a kind of lexicographic order $\Rightarrow$, where $(i, j) \Rightarrow (k,l)$ if and only if $j > l$ or both $j = l$ and $i \geq k$. Otherwise, $(i, j) \rightarrow (k,l)$ means $j > l$ or both $j = l$ and $i > k$. $(i, j) \geq (k,l)$ implies that $(i, j) \Rightarrow (k, l)$.
	
	\paragraph*{} A nonzero bivariate polynomial $f(x, y) \in \mathbb{F}_q[x, y; \rho, \theta]$ is said to have quasi-degree $deg(f(x, y)) = (k, \lambda)$ if $f(x, y)$ has a nonzero term $a_{k,\lambda}x^k y^\lambda$ but not any nonzero term $a_{i,j}x^i y^j$ such that $(i, j) \rightarrow (k, \lambda)$ holds. $deg(f_1(x, y)) = (k_1, \lambda_1) \geq deg(f_2(x, y)) = (k_2, \lambda_2)$ means $k_1 \geq  k_2$ and $\lambda_1 \geq \lambda_2$.
	
	\paragraph*{} Then for any polynomials $f(x, y)$ and $g(x, y)$ in $\mathbb{F}_q[x, y; \rho, \theta]$, we have
	\begin{enumerate}
		\item $deg(f(x, y) + g(x, y)) \Leftarrow \max\{deg(f(x, y)), deg(g(x, y))\}$, where max is under the lexicographic order.
		\item $deg(f(x, y) \ast g(x, y)) = deg(f(x, y)) + deg(g(x, y))$.
	\end{enumerate}
	
	\paragraph*{} The following facts are straightforward for $\mathbb{F}_q[x, y; \rho, \theta]$.
	\begin{enumerate}
		\item It has no nonzero zero-divisors.
		\item The units of $\mathbb{F}_q[x, y; \rho, \theta]$ are the units of $\mathbb{F}_q$.
		\item The commutative law does not hold.
	\end{enumerate}
	
	\begin{thm}\textup{(See Li and Li \cite{li20142-D})}
		Let $f_1(x, y), f_2(x, y) \in \mathbb{F}_q[x, y; \rho, \theta]$ be two nonzero bivariate polynomials. Provided that $deg(f_1(x, y))  \geq deg(f_2(x, y))$, there exists a pair of polynomials $h(x, y)(\neq 0), ~g(x, y) \in \mathbb{F}_q[x, y; \rho, \theta]$ which satisfy
		\begin{center}
			$f_1(x, y) = h(x, y) \ast f_2(x, y) + g(x, y)$
		\end{center}
		such that $g(x, y) = 0$ or $deg(f_2(x, y)) \nleq deg(g(x, y)) (\leftarrow deg(f_1(x, y)))$.\smallskip\\
		We call this \em{right division} of $f_1(x, y)$ by $f_2(x, y)$.
	\end{thm}
	
	\begin{thm}\textup{(See Li and Li \cite{li20142-D})}
		Suppose that $|\langle \rho \rangle| \, | \, s, |\langle \theta \rangle| \, | \, l$. Then for any $f(x, y) \in \mathbb{F}_q[x, y; \rho, \theta], (x^s - 1) \ast f(x, y) = f (x, y) \ast (x^s - 1)$ and $(y^l - 1) \ast f(x, y) = f (x, y) \ast (y^l - 1)$.
	\end{thm}
	
	\paragraph*{} So when $|\langle \rho \rangle| \, | \, s, |\langle \theta \rangle| \, | \, l$, the left ideal $\langle x^s-1, y^l-1\rangle_l$ is indeed an ideal in $\mathbb{F}_q[x, y; \rho, \theta]$.
	
	\begin{defn}\textup{(See Li and Li \cite{li20142-D})}
		\textup{Let $\theta$ be an automorphism of $\mathbb{F}_q$. A subset $C$ of $\mathbb{F}^n_q$ is called a \em{skew-quasi-cyclic} code of length $n$ and index $l$ under $\theta$ (or $l_\theta$-SQC code) where $n = sl$ if\medskip\\
			(1) $C$ is a subspace of $\mathbb{F}^n_q$;\smallskip\\
			(2) if $c = (c_{0,0}, \dots, c_{0,l-1}, c_{1,0}, \dots, c_{1,l-1}, \dots, c_{s-1,0}, \dots, c_{s-1,l-1})$ is a codeword of $C$, then $(\theta(c_{s-1,0}), \dots, \theta(c_{s-1,l-1}), \theta(c_{0,0}), \dots, \theta(c_{0,l-1}), \dots, \theta(c_{s-2,0}), \dots, \theta(c_{s-2,l-1}))$ is also a codeword in $C$.}
	\end{defn}
	
	\paragraph*{} Let $C$ be a linear code over $\mathbb{F}_q$ of length $sl$ whose codewords are viewed as $s \times l$ arrays, i.e. $c \in C$ is written as
	$$c =	
	\begin{pmatrix}
		c_{0,0} & \cdots & c_{0,l-1}\\
		c_{1,0} & \cdots & c_{1,l-1}\\
		& \vdots &			\\
		c_{s-1,0} & \cdots & c_{s-1,l-1}
	\end{pmatrix}.$$
	
	\paragraph*{} Let
	$$C_1 = \left\{(c_{0,0}, \dots, c_{0,l-1}, \dots, c_{s-1,0}, \dots, c_{s-1,l-1}) : \begin{pmatrix}
		c_{0,0} & \cdots & c_{0,l-1}\\
		c_{1,0} & \cdots & c_{1,l-1}\\
		& \vdots &			\\
		c_{s-1,0} & \cdots & c_{s-1,l-1}
	\end{pmatrix} \in C\right\},$$
	$$C_2 = \left\{(c_{0,0}, \dots, c_{s-1,0}, \dots, c_{0,l-1}, \dots, c_{s-1,l-1}) : \begin{pmatrix}
		c_{0,0} & \cdots & c_{0,l-1}\\
		c_{1,0} & \cdots & c_{1,l-1}\\
		& \vdots &			\\
		c_{s-1,0} & \cdots & c_{s-1,l-1}
	\end{pmatrix} \in C\right\}.$$
	
	\paragraph*{} From the definition we know that $C$ is a 2-D skew-cyclic code if and only if $C_1$ is a skew-quasi-cyclic code of length $sl$ and index $l$ under $\rho$ and $C_2$ is a skew-quasi-cyclic code of length $sl$ and index $s$ under $\theta$.
	
	\paragraph*{} Let $R = \mathbb{F}_q[x, y; \rho, \theta]/\langle x^s-1, y^l-1\rangle_l$ denote the set of all polynomials over $\mathbb{F}_q$ of degree less than $s$ with respect to $x$ and of degree less than $l$ with respect to $y$.
	
	\paragraph*{} Consider the following $\mathbb{F}_q$-space isomorphism
	\begin{align*}
		\mathbb{F}^{s \times l}_q & \rightarrow R,\medskip\\
		(a_{ij}) & \rightarrow \sum\limits_{i=0}^{s-1} \sum\limits_{j=0}^{l-1}a_{ij}x^iy^j,
	\end{align*}
	where $\mathbb{F}^{s \times l}_q$ denotes the space of all $s \times l$ arrays. Then a codeword $c \in C$ can be denoted by bivariate polynomial $c(x, y)$ under the above isomorphism.
	
	\paragraph*{} Furthermore, for $f(x, y) \in \mathbb{F}_q[x, y; \rho, \theta]$ let $\{f(x, y)\}_R$ denote the polynomial in $R$ such that
	\begin{center}
		$\{f(x, y)\}_R - f(x, y) \in \langle x^s - 1, y^l - 1\rangle_l,$
	\end{center}
	where $\langle x^s - 1, y^l - 1\rangle_l$ denotes the left ideal generated by $x^s - 1$ and $y^l - 1$ in $\mathbb{F}_q[x, y; \rho, \theta]$.
	
	\paragraph*{} For an arbitrary bivariate polynomial $f(x, y)$ over $\mathbb{F}_q$. We have
	\begin{center}
		$f(x, y) = \{f(x, y)\}_R + a(x, y) \ast (x^s - 1) + b(x, y) \ast (y^l - 1)$,
	\end{center}
	where $a(x, y)$ and $b(x, y)$ are bivariate polynomials over $\mathbb{F}_q$.
	
	\paragraph*{} Let $\{f(x, y)\}_R$ be an element in $R$, and let $r(x, y) \in \mathbb{F}_q[x, y; \rho, \theta]$. Define multiplication from left as:
	\begin{center}
		$r(x, y) \ast \{f(x, y)\}_R = \{r(x, y) \ast f (x, y)\}_R.$
	\end{center}
	This is a well-defined multiplication of the elements of $R$ by the elements of $\mathbb{F}_q[x, y; \rho, \theta]$ from the left. Under this definition, $R$ is a left $\mathbb{F}_q[x, y; \rho, \theta]$-module.
	
	\paragraph*{} Using the polynomial representation, a 2-D skew-cyclic code is defined as a two-dimensional linear code such that for each codeword $c(x, y), \{x \ast c(x, y)\}_R$, and $\{y \ast c(x, y)\}_R$ are also codewords. Thus we conclude that a linear subspace of $\mathbb{F}_q^{s \times l}$ is 2-D skew-cyclic if and only if $C$ is a left submodule when viewed as a subset of $R$.
	
	\begin{thm}\textup{(See Li and Li \cite{li20142-D})}
		A code $C$ in $R$ is a $2$-D skew-cyclic code if and only if $C$ is a left $\mathbb{F}_q[x, y; \rho, \theta]$-submodule of the left $\mathbb{F}_q[x, y; \rho, \theta]$-module $R$.
	\end{thm}
	
	\paragraph*{} Let $D \supseteq \langle x^s - 1, y^l - 1\rangle_l$ be a left ideal of $\mathbb{F}_q[x, y; \rho, \theta]$, then $D/\langle x^s - 1, y^l - 1\rangle_l$ is a left $\mathbb{F}_q[x, y; \rho, \theta]$-submodule of $R$. That is to say $D/\langle x^s - 1, y^l - 1\rangle_l$ produce a 2-D skew-cyclic code.
	
	\paragraph*{} Suppose that $D = \langle f_1, f_2, \dots, f_m\rangle_l$, then $\langle \{f_1\}_R, \{f_2\}_R, \dots, \{f_m\}_R \rangle_l$ produce a 2-D skew-cyclic code. Recall for any ring $A$, we denote the center of $A$ by $Z(A)$.
	
	\begin{lemma}\textup{(See Li and Li \cite{li20142-D})}
		$x^s - 1 \in Z(\mathbb{F}_q[x, y; \rho, \theta])$ if and only if $|\langle \rho \rangle| \, | \, s$. $y^l-1 \in Z(\mathbb{F}_q[x, y; \rho, \theta])$ if and only if $|\langle \theta \rangle| \, | \, l$.
	\end{lemma}
	
	\begin{thm}\textup{(See Li and Li \cite{li20142-D})}
		Let $|\langle \rho \rangle| \, | \, s, |\langle \theta \rangle| \, | \, l$. A code $C$ in $R$ is a $2$-D skew-cyclic code if and only if $C$ is a left ideal of $R$.
	\end{thm}

	\begin{thm}\textup{(See Li and Li \cite{li20142-D})}
		Let $C$ be a $2$-D skew-cyclic code over $\mathbb{F}_q$ of size $s \times l$ under $\rho$ and $\theta$. If $(s,l) = 1$, then $C$ is equivalent to a skew-cyclic code.
	\end{thm}
	
	\begin{coro}\textup{(See Li and Li \cite{li20142-D})}
		Let $C$ be a $2$-D skew-cyclic code over $\mathbb{F}_q$ of size $s \times l$ under $\rho$ and $\theta$. If $(s,l) = 1$ and $\rho \theta$ is the identity mapping, then $C$ is equivalent to a cyclic code.
	\end{coro}

	\subsection{Generator matrix for 2-D skew-cyclic codes over $\mathbb{F}_q [x, y; \rho, \theta]$}
	
	\paragraph*{} Theorem 6.7 suggests that the study of left ideals of $R$ is equivalent to studying 2-D skew-cyclic codes. This idea was used by Sepasdar in \cite{sepasdar2016some} to find the generator matrix of 2-D skew-cyclic codes, this work is an extension of the ideas presented by the author in \cite{sepasdar2017generator}.
	
	\paragraph*{} Let $I$ is an ideal of the ring $R \coloneqq \mathbb{F}_q[x, y; \rho, \theta]/\langle x^s-1, y^l-1\rangle_l$. Thus we can write every element $f(x, y) \in I$ uniquely as
	$f(x, y) = \sum_{i = 0}^{l - 1}f_i(x)y^i$, where each $f_i(x) \in R_1 \coloneqq \mathbb{F}_q[x; \theta]/\langle x^s-1\rangle$. Observing that the ring $R_1$ is a principal ideal ring, any left ideal of $R_1$ can be generated by a right divisor of $x^s - 1$ in $\mathbb{F}_q[x, y; \rho, \theta]$.
	
	\paragraph*{} Take$I_0 = \{g_0(x) \in R_1 :$ $\exists$ $g(x, y) \in I$ such that $g(x, y) = \sum_{i = 0}^{l - 1}g_i(x)y^i\}$.
	
	\paragraph*{} Thus $I_0$ is a left ideal of $R_1$, so $\exists$ a unique monic polynomial $p_0^0(x) \in R_1$ such that $I_0 = \langle p_0^0(x) \rangle$ and $\exists \, q_0(x)$
	such that $x^s - 1 = q_0(x) p_0^0(x)$. Now, since $f(x, y) \in I$ it is clear that
	$f_0(x) \in I_0$, and so $f_0(x) = p_0'(x) p_0^0(x)$ for some $p_0'(x) \in R_1$. On the other hand, $p_0^0(x) \in I_0$ implies that $\exists \, \mathbf{p}_0(x, y) \in I$ such that $\mathbf{p}_0(x, y) = \sum_{i = 0}^{l - 1}p^0_i(x)y^i$. Set $h_1(x, y) = f(x, y) - p_0'(x) \mathbf{p}_0(x, y)$, hence
	$h_1(x, y) \in I$ and is of the form $h_1(x, y) = \sum_{i = 1}^{l - 1}h^1_i(x)y^i$.
	
	\paragraph*{} Take $I_1 = \{g_1(x) \in R_1 : \, \exists \, g(x, y) \in I$ such that $g(x, y) = \sum_{i = 1}^{l - 1}g_i(x)y^i\}$.
	
	\paragraph*{} Clearly $I_1$ is a left ideal of $R_1$, so $\exists$ unique monic polynomial $p_1^1(x) \in R_1$ such that $I_1 = \langle p_1^1(x) \rangle$ and $\exists \, q_1(x)$
	such that $x^s - 1 = q_1(x) p_1^1(x)$. Now, since $h_1(x, y) \in I$ it is clear that
	$h_1^1(x) \in I_1$, and so $h_1^1(x) = p_1'(x) p_1^1(x)$ for some $p_1'(x) \in R_1$. On the other hand, $p_1^1(x) \in I_1$ implies that $\exists \, \mathbf{p}_1(x, y) \in I$ such that $\mathbf{p}_1(x, y) = \sum_{i = 1}^{l - 1}p^1_i(x)y^i$. Set $h_2(x, y) = h_1(x, y) - p_1'(x) \mathbf{p}_1(x, y)$, hence
	$h_2(x, y) \in I$ and is of the form $h_2(x, y) = \sum_{i = 2}^{l - 1}h^2_i(x)y^i$.
	
	\paragraph*{} Take $I_2 = \{g_2(x) \in R_1 : \, \exists \,  g(x, y) \in I$ such that $g(x, y) = \sum_{i = 2}^{l - 1}g_i(x)y^i\}$.
	
	\paragraph*{} By this method one can construct $\mathbf{p}_i(x, y) \in I$ for $i = 0, \dots, l - 1$. It is easy to see that
	\begin{center}
		$I = \langle \mathbf{p}_0(x, y), \dots, \mathbf{p}_{l - 1}(x, y) \rangle$.
	\end{center}
	We call $\mathbf{p}_i(x, y)$ as the generating polynomials of $I$.
	
	\paragraph*{} By multiplying $y^j \, (j = 1, \dots, l - 1)$ to each $\mathbf{p}_i(x, y) \, \, (i = 1, \dots, l - 1)$ we can find some conditions on the terms of generating polynomials which are provided in \cite{sepasdar2016some}. By the following theorem in \cite{sepasdar2016some}, one can easily find the generator matrix and dimension of two dimensional skew cyclic codes.
	
	\begin{thm}\textup{(See Sepasdar \cite{sepasdar2016some})}
		Let $I$ be a left ideal of the ring
		$R = \mathbb{F}_q[x, y; \rho, \theta]/\langle x^s-1, y^l-1\rangle_l$  (a two dimensional skew cyclic code) with length $n = sl$ and is generated by $\{\mathbf{p}_0(x, y), \dots, \mathbf{p}_{l - 1}(x, y)\}$, which is obtained from the first method. Then the set
		\begin{align*}
			& \{\mathbf{p}_0(x, y), x\mathbf{p}_0(x, y), \dots, x^{n - a_0 - 1}\mathbf{p}_0(x, y),\medskip\\
			& \mathbf{p}_1(x, y), x\mathbf{p}_1(x, y), \dots, x^{n - a_1 - 1}\mathbf{p}_1(x, y),\medskip\\
			& \hspace{3.5cm} \vdots\medskip\\
			& \mathbf{p}_{l - 1}(x, y), x\mathbf{p}_{l - 1}(x, y), \dots, x^{n - a_{l - 1} - 1}\mathbf{p}_{l - 1}(x, y)\}
		\end{align*}
		where $a_i \coloneqq \deg(p^i_i(x))$ for $i = 0, \dots, l - 1$, is an $\mathbb{F}_q$-basis for $I$.
	\end{thm}
	
	\subsection{2D skew-cyclic codes over $\mathbb{F}_q + u\mathbb{F}_q$}
	
	\paragraph*{} The paper \cite{sharma2019a class} by Sharma and Bhaintwal presented 2D skew-cyclic codes over the ring $R = \mathbb{F}_q + u\mathbb{F}_q, u^2 = 1$. $R$ is observed to be a quotient ring $\mathbb{F}_q[u]/\langle u^2 - 1 \rangle$ and forms a semi-local ring with two maximal ideals $\langle 1 + u \rangle$ and $\langle 1 - u \rangle$.
	
	\paragraph*{} A Gray map $\phi : R \rightarrow \mathbb{F}_q^2$ is defined  as follows
	\begin{center}
		$\phi(a + ub) = (b, a + b)$,
	\end{center}
	which can be extended componentwise to a linear map $\phi : R^n \rightarrow \mathbb{F}_q^{2n}$. Further, we define the Gray weight $w_G(x)$ of any $x \in R^n$ as $w_G(x) = w_H(\phi(x))$, where $w_H(\phi(x))$ denote the Hamming weight of $\phi(x)$.
	
	\paragraph*{} The authors considered a finite commutative ring $R$ with identity along with two of its automorphisms $\theta_1, \theta_2$ imposing the additional constraint that $\theta_1 \theta_2 = \theta_2 \theta_1$. On the same lines as \cite{li20142-D} they considered the bivariate skew polynomial ring $R[x, y, \theta_1, \theta_2]$, i.e.,
	\begin{center}
		$R[x, y, \theta_1, \theta_2] = \left\{ \sum\limits_{i = 0}^{l - 1} \sum\limits_{j = 0}^{m - 1} a_{i, j}x^iy^j \, : \, a_{i, j} \in R; \, m,l \in \mathbb{N}\right\}$,
	\end{center}
	with usual addition of polynomials. But as $\theta_1$ and $\theta_2$ are now commutative the  multiplication rule becomes
	\begin{center}
		$x^iy^ja = \theta_1^i \theta_2^j(a)x^iy^j$.
	\end{center}
	
	\paragraph*{} Again, using a lexicographic ordering on the polynomials of $R[x, y, \theta_1, \theta_2]$ they provide a generalization for the results in \cite{li20142-D, sepasdar2016some, sepasdar2017generator} for the case $\theta_1 \theta_2 = \theta_2 \theta_1$ by showing that generator polynomials of $2$D skew-cyclic codes are precisely divisors of $(x^l - 1)(y^m - 1)$.
	
	\begin{thm}\textup{(See Sharma and Bhaintwal \cite{sharma2019a class})}
		Let $C$ be a $2$D skew-cyclic code of length $ml$ over $R$. If the order of $\theta_1$
		divides $l$ and the order of $\theta_2$ divides $m$, then the following hold.
		\begin{enumerate}[(1)]
			\item The polynomials $x^l - 1, y^m - 1 \in R[x, y, \theta_1, \theta_2]$ are in $Z (R[x, y, \theta_1, \theta_2])$, where
			$Z(R[x, y, \theta_1, \theta_2])$ is the center of $R[x, y, \theta_1, \theta_2]$.
			\item $C$ is an ideal of $R[x, y, \theta_1, \theta_2]/\langle x^l - 1, y^m - 1\rangle$.
			\item If $f(x, y) \in Z(R[x, y, \theta_1, \theta_2])$ is a monic polynomial which factorizes into
			a product of two monic polynomials as $h(x, y)g(x, y)$ in $Z(R[x, y, \theta_1, \theta_2])$, then
			$h(x, y)g(x, y) = g(x, y)h(x, y)$. In particular, any right divisor of $(x^l - 1)(y^m - 1)$
			is also a left divisor thereof.
		\end{enumerate}
	\end{thm}
	
	\paragraph*{} The main goal of the paper was to study a class of 2D skew-cyclic codes over $R = \mathbb{F}_q + u\mathbb{F}_q, u^2 = 1$,
	using a bivariate polynomial ring over $R$. The authors considered two types of
	automorphisms of $R$ namely, $\theta$ and $\sigma_i$ defined as
	$\theta(a + ub) = a - ub$,
	and $\sigma_i (a + ub) = a^{p^i} + ub^{p^i}, \, i \leq r, \, i \, | \, r$, respectively. These special type of automorphisms of $R$ have the property that $\theta \sigma_i = \sigma_i \theta$ for all $i$.
	
	\paragraph*{} For simplicity, the notation $E = \{\sigma_i \, : \, i \leq r, \, i \, | \, r\}$ is used. Also we have $|\sigma_i| = r/i$. Further, assume $\theta_1 = \theta$ and $\theta_2 = \sigma$, where $\sigma \in E$.
	
	\paragraph*{} Consider $M$ to be a left submodule of $R_{m, l} = R[x, y, \theta, \sigma]/\langle x^l - 1, y^m - 1\rangle$ . Then a minimal set $B = \{f_1, f_2, \dots, f_s\} \subseteq M$ is called a consistent set of $M$ if it satisfies the following
	conditions:
	\begin{enumerate}[(1)]
		\item $f_i \ngeq f_j, \, 1 \leq i, \, j \leq s, \, i \neq j$, if the lex-leading coefficients of $f_i$ and $f_j$ are
		either both unit or both non-units.
		\item for any $f \in M$ such that the lex-leading coefficient of $f$ is a unit, there exists
		some $f_i \in B, \, 1 \leq i \leq s$, with its leading coefficient a unit such that $f \geq f_i$.
		\item for any $f \in M$ such that the lex-leading coefficient of $f$ is a non-unit, there exists
		some $f_j \in B, \, 1 \leq j \leq s$, such that $f \geq f_j$.
	\end{enumerate}

	\paragraph*{} The next few theorems from \cite{sharma2019a class} determine the relationships between 2D skew-cyclic codes over $R$ and consistent sets of $M$.
	
	\begin{thm}\textup{(See Sharma and Bhaintwal \cite{sharma2019a class})}
		Let $C$ be a $2$D skew-cyclic code of length $ml$ over $R$. If $C$ has a consistent
		set $B = \{f_1, f_2, \dots, f_s\}$ such that the lex-leading coefficient of each $f_i$ is a unit, then
		$C = \langle f_1, f_2, \dots, f_s\rangle$.
	\end{thm}

	\begin{thm}\textup{(See Sharma and Bhaintwal \cite{sharma2019a class})}
		Let $C$ be a $2$D skew-cyclic code of length $ml$ over $R$. If $C$ contains a polynomial $g(x, y)$ with its lex-leading coefficient a unit such that $g(x, y) \leq c(x, y)$ for all $c(x, y) \in C$, then $C = \langle g(x, y) \rangle$ and $g(x, y) \, | \, (x^l - 1)(y^m - 1)$. Moreover,
		the set
		\begin{center}
			$S = \left\{\begin{array}{cccc}
				g(x, y), & xg(x, y), & \cdots & ,x^{l - a - 1}g(x, y)\\
				yg(x, y), & yxg(x, y), & \cdots & ,yx^{l - a - 1}g(x, y)\\
				\vdots & \vdots & \ddots & \vdots\\
				y^{m - b - 1}g(x, y), & y^{m - b - 1}xg(x, y), & \cdots & ,y^{m - b - 1}x^{l - a - 1}g(x, y)
			\end{array}\right\}$
		\end{center}
		is a basis for $C$, and hence $|C| = |R|^{(l - a)(m - b)}$, where $(a, b) =$ lexdeg $g(x, y)$.
	\end{thm}
	
	\begin{thm}\textup{(See Sharma and Bhaintwal \cite{sharma2019a class})}
		Let $C$ be a $2$D skew-cyclic code of length $ml$ in $R[x, y, \theta, \sigma]$ such that it does not contain any polynomial with its lex-leading coefficient a unit. Let $B = \{f_1, f_2, \dots, f_m\}$ be a consistent set of $C$. Then $C = \langle f_1, f_2, \dots, f_m \rangle$.
	\end{thm}

	\paragraph*{} Following results from \cite{sharma2019a class} state the various relationships between 2D skew-cyclic codes, 2D cyclic codes, skew cyclic codes and cyclic codes over the ring $R = \mathbb{F}_q + u\mathbb{F}_q$.
	\begin{thm}\textup{(See Sharma and Bhaintwal \cite{sharma2019a class})}
		Let $C$ be a $2$D skew-cyclic code of length $ml$ over $R$ with the corresponding
		automorphisms as $\theta$ and $\sigma$ such that $\gcd(l, |\theta|) = 1$ and $\gcd(m, |\sigma|) = 1$. Then $C$ is a $2$D cyclic code of the same length over $R$.
	\end{thm}

	\begin{thm}\textup{(See Sharma and Bhaintwal \cite{sharma2019a class})}
		Let $C$ be a $2$D skew-cyclic code of length $n = ml$ over $R$. If $\gcd(m, l) = 1$, then $C$ is equivalent to a skew-cyclic code of length $n$ over $R$.
	\end{thm}

	\begin{coro}\textup{(See Sharma and Bhaintwal \cite{sharma2019a class})}
		Let $C$ be a $2$D skew-cyclic code of length $n = ml$ over $R$. If $\gcd(m, l) = 1$ and $\theta \sigma$ is the identity map on $R$, then $C$ is equivalent to a cyclic code of the same length over $R$.
	\end{coro}

	\paragraph*{} Next, the authors study the duals of 2D skew-cyclic codes of length $n = ml$ over $R$
	when $|\theta| \, | \, l$ and $|\sigma| \, | \, m$.
	
	\begin{thm}\textup{(See Sharma and Bhaintwal \cite{sharma2019a class})}
		Let $C$ be a $2$D skew-cyclic code of length $n = ml$ over $R$ with the associated automorphisms $\theta$ and $\sigma$ such that $|\theta| \, | \, l$ and $|\sigma| \, | \, m$. Then $C^\perp$ is also a $2$D skew-cyclic code of the same length over $R$.
	\end{thm}

	\paragraph*{} If $p$ to be an odd prime, the decomposition of a 2D
	skew-cyclic code over $R$ into 2D skew-cyclic codes over $\mathbb{F}_q$ is provided in the following theorem:
	
	\begin{thm}\textup{(See Sharma and Bhaintwal \cite{sharma2019a class})}
		Let $\sigma', \sigma'' \in E$ be two automorphisms of $R$. Let $C_{\sigma', \sigma''} = (1 + u)C' \oplus (1 - u)C''$ be a linear code over $R$. Then $C$ is a $2$D skew-cyclic code of length $n$ over $R$
		if and only if $C'$ and $C''$ are $2$D skew-cyclic codes over $\mathbb{F}_q$ in which the corresponding
		automorphisms are $\sigma', \sigma''$, restricted to $\mathbb{F}_q$.
	\end{thm}

	\section{2-D skew constacyclic codes over arbitrary commutative ring}
	
	\paragraph*{} Mostafanasab introduced the 2-D skew constacyclic codes over $R[x, y; \rho, \theta]$ in \cite{mostafanasab20152-dskew}. These codes are generalizations of 2-D skew-cyclic codes over $\mathbb{F}_q[x, y; \rho, \theta]$ which we encountered earlier.
	
	\paragraph*{} For this we consider $R$ to be a commutative ring together with two of its automorphisms $\rho$ and $\theta$ and the set of formal bivariate polynomials
	\begin{center}
		$R[x, y; \rho, \theta] = \{\sum \limits_{j = 0}^{t} \sum \limits_{i = 0}^{k} a_{i,j}x^iy^j \, | \, a_{i,j} \in R$ and $k, t \in \mathbb{N}\}.$
	\end{center}
	which forms a ring under the usual polynomial addition and the following multiplication rule
	\begin{center}
		$ax^iy^j \star bx^ry^s = a\rho^i\theta^j(b)x^{i+r}y^{j+s}$,
	\end{center}
    extended to all elements of $R[x, y; \rho, \theta]$ by associativity and distributivity. Unless $\rho$ and $\theta$ are identity automorphisms, $R[x, y; \rho, \theta]$ over $R$ is a non-commutative ring. Let the left
	ideal of $R[x, y; \rho, \theta]$ generated by $f(x, y)$ is written as $\langle f(x, y) \rangle_l$, it need not be two-sided.  Also, denote by $R^{\rho, \theta}$ (resp. $R^{\rho \theta}$) the subring of $R$ that is fixed by $\rho, \theta$ (resp. $\rho \theta$).
	
	\begin{thm}\textup{(See Mostafanasab \cite{mostafanasab20152-dskew})}
		Let $f(x, y) \in R^{\rho, \theta}[x, y; \rho, \theta]$. If $f(x, y) \star g(x, y) = 0$ for
		some $0 \neq g(x, y) \in R[x, y; \rho, \theta]$, then there exists $0 \neq r \in R$ such that
		$f(x, y) \star r = 0$.
	\end{thm}

	\paragraph*{} We use the notation $Z(S)$ to denote the center of the ring $S$. Next few results taken from \cite{mostafanasab20152-dskew} provide necessary and sufficient conditions for an ideal to be two-sided in $R[x, y; \rho, \theta]$.
	\begin{prop}\textup{(See Mostafanasab \cite{mostafanasab20152-dskew})}
		Let $g(x, y) \in Z(R[x, y; \rho, \theta])$. Then $\langle x^ky^t \star g(x, y)\rangle_l$ is a two-sided ideal in $R[x, y; \rho, \theta]$ for every $k, t \in \mathbb{N}$.
	\end{prop}
	
	\begin{prop}\textup{(See Mostafanasab \cite{mostafanasab20152-dskew})}
		Let $|\langle \rho \rangle | \, | \, l$ and $|\langle \theta \rangle | \, | \, s$. Then $R^{\rho, \theta}[x^l, y^s; \rho, \theta]  \subseteq Z(R[x, y; \rho, \theta])$.
	\end{prop}

	\begin{prop}\textup{(See Mostafanasab \cite{mostafanasab20152-dskew})}
		Let $\lambda$ be a unit in $R$. The following conditions hold:
		\begin{enumerate}[(1)]
			\item If $\langle x^l - \lambda \rangle_l$ is a two-sided ideal of $R[x, y; \rho, \theta]$, then $|\langle \rho \rangle | \, | \, l$ and $\rho(\lambda) = \lambda$.
			\item If $\langle y^s - \lambda \rangle_l$ is a two-sided ideal of $R[x, y; \rho, \theta]$, then $|\langle \theta \rangle | \, | \, s$ and $\theta(\lambda) = \lambda$.
		\end{enumerate}
	\end{prop}
	
	\begin{prop}\textup{(See Mostafanasab \cite{mostafanasab20152-dskew})}
		Let $\lambda_1, \lambda_2$ be a units in $R$. The following conditions hold:
		\begin{enumerate}[(1)]
			\item If $|\langle \theta \rangle | \, | \, s$ and $\langle x^l - \lambda_1, y^s - \lambda_2 \rangle_l$ is a two-sided ideal of $R[x, y; \rho, \theta]$, then $|\langle \rho \rangle | \, | \, l$ and $\rho(\lambda_1) = \lambda_1$.
			\item If $|\langle \rho \rangle | \, | \, l$ and $\langle x^l - \lambda_1, y^s - \lambda_2 \rangle_l$ is a two-sided ideal of $R[x, y; \rho, \theta]$, then $|\langle \theta \rangle | \, | \, s$ and $\theta(\lambda_2) = \lambda_2$.
		\end{enumerate}
	\end{prop}

		\begin{thm}\textup{(See Mostafanasab \cite{mostafanasab20152-dskew})}
		Let $\lambda \in R$. The following conditions hold:
		\begin{enumerate}[(1)]
			\item If $|\langle \rho \rangle | \, | \, l$ and $\lambda \in R^{\rho, \theta}$, then $x^l - \lambda$ is central in $R[x, y; \rho, \theta]$.
			\item If $|\langle \theta \rangle | \, | \, s$ and $\lambda \in R^{\rho, \theta}$, then $y^s - \lambda$ is central in $R[x, y; \rho, \theta]$.
			\item If $|\langle \rho \rangle | \, | \, l, |\langle \theta \rangle | \, | \, s$ and $\lambda_1, \lambda_2 \in R^{\rho, \theta}$, then $\langle x^l - \lambda_1, y^s - \lambda_2 \rangle_l$ is a two-sided ideal of $R[x, y; \rho, \theta]$.
		\end{enumerate}
	\end{thm}

	\begin{coro}\textup{(See Mostafanasab \cite{mostafanasab20152-dskew})}
		Let $R$ be a ring. The following conditions hold:
		\begin{enumerate}[(1)]
			\item If $\theta(\lambda) = \lambda$. Then $|\langle \rho \rangle | \, | \, l$ and $\rho(\lambda) = \lambda$ if and only if $x^l - \lambda$ is central in $R[x, y; \rho, \theta]$.
			\item If $\rho(\lambda) = \lambda$. Then $|\langle \theta \rangle | \, | \, s$ and $\theta(\lambda) = \lambda$ if and only if $y^s - \lambda$ is central in $R[x, y; \rho, \theta]$.
		\end{enumerate}
	\end{coro}

	\begin{prop}\textup{(See Mostafanasab \cite{mostafanasab20152-dskew})}
		Let $f \star g$ be a monic central bivariat skew polynomial in $R[x, y; \rho, \theta]$ for some $f, g \in R[x, y; \rho, \theta]$. Then $f \star g = g \star f$.
	\end{prop}

	\begin{coro}\textup{(See Mostafanasab \cite{mostafanasab20152-dskew})}
		Let $|\langle \rho \rangle | \, | \, l, |\langle \theta \rangle | \, | \, s$ and $\lambda_1, \lambda_2 \in R^{\rho, \theta}$. If $(x^l - \lambda_1) \star (y^s - \lambda_2) = f \star g$ for some $f, g \in R[x, y; \rho, \theta]$, then $f \star g = g \star f$.
	\end{coro}

	\paragraph*{} Assume $C$ to be a linear code of length $ls$ over $R$. The codewords of $C$ are canb be represented as $l \times s$ arrays, i.e., $c \in C$ has the form
	\begin{center}
		$\begin{pmatrix}
			c_{0, 0} & c_{0, 1} & \cdots & c_{0, s - 1}\\
			c_{1, 0} & c_{1, 1} & \cdots & c_{1, s - 1}\\
			\vdots & \vdots & \ddots & \vdots\\
			c_{l - 1, 0} & c_{l - 1, 1} & \cdots & c_{l - 1, s - 1}
		\end{pmatrix}$.
	\end{center}

	\paragraph*{} $C$ is said to be a column skew $\lambda_1$-constacyclic code if for every $l \times s$
	array $c = (c_{i, j}) \in C$ we have that
	\begin{center}
		$\begin{pmatrix}
			\lambda_1\rho(c_{l - 1, 0}) & \lambda_1\rho(c_{l - 1, 1}) & \cdots & \lambda_1\rho(c_{l - 1, s - 1})\\
			\rho(c_{0, 0}) & \rho(c_{0, 1}) & \cdots & \rho(c_{0, s - 1})\\
			\vdots & \vdots & \ddots & \vdots\\
			\rho(c_{l - 2, 0}) & \rho(c_{l - 2, 1}) & \cdots & \rho(c_{l - 2, s - 1})
		\end{pmatrix} \in C$.
	\end{center}

	\paragraph*{} Also, it is a row skew $\lambda_2$-constacyclic code if for every $l \times s$
	array $c = (c_{i, j}) \in C$ we have that
	\begin{center}
		$\begin{pmatrix}
			\lambda_2\theta(c_{0, s - 1}) & \theta(c_{0, 0}) & \cdots & \theta(c_{0, s - 2})\\
			\lambda_2\theta(c_{1, s - 1}) & \theta(c_{1, 0}) & \cdots & \theta(c_{1, s - 2})\\
			\vdots & \vdots & \ddots & \vdots\\
			\lambda_2\theta(c_{l - 1, s - 1}) & \theta(c_{l - 1, 0}) & \cdots & \theta(c_{l - 1, s - 2})
		\end{pmatrix} \in C$.
	\end{center}
	
	\paragraph*{} $C$ is called a 2-D skew $(\lambda_1, \lambda_2)$-constacyclic code if it is both column skew $\lambda_1$-constacyclic and row skew $\lambda_2$-constacyclic.
	
	\paragraph*{} From the earlier discussion, once we have two-sided ideals, we can define the quotient ring $R^\circ \coloneqq R[x, y; \rho, \theta]/\langle x^l - \lambda_1, y^s - \lambda_2\rangle_l$. Let $R^{l \times s}$ denotes the set of all $l \times s$ arrays and consider the $R$-module isomorphism $R^{l \times s} \rightarrow R^\circ$ defined by
	\begin{center}
		$(a_{i, j}) \mapsto \sum\limits_{j = 0}^{s - 1} \sum\limits_{i = 0}^{l - 1} a_{i, j}x^iy^j$.
	\end{center}
	Under the above isomorphism for every codeword $c \in C$ we have its skew polynomial representation $c(x, y)$ as we earlier saw for 2-D skew cyclic codes.
	
	\paragraph*{} The next few results show that $2$-D skew $(\lambda_1, \lambda_2)$-constacyclic codes are precisely left ideals of $R^\circ$.
	
	\begin{thm}\textup{(See Mostafanasab \cite{mostafanasab20152-dskew})}
		A code $C$ in $R^\circ$ is a $2$-D skew $(\lambda_1, \lambda_2)$-constacyclic code if and only if $C$ is a left $R[x, y; \rho, \theta]$-submodule of the left $R[x, y; \rho, \theta]-$module $R^\circ$.
	\end{thm}

	\begin{coro}\textup{(See Mostafanasab \cite{mostafanasab20152-dskew})}
		Let $|\langle \rho \rangle | \, | \, l, |\langle \theta \rangle | \, | \, s$ and $\lambda_1, \lambda_2 \in R^{\rho, \theta}$. A code $C$ in $R^\circ$ is a $2$-D skew $(\lambda_1, \lambda_2)$-constacyclic code if and only if $C$ is a left ideal of $R^\circ$.
	\end{coro}

	\paragraph*{} A series of theorems and lemmas from \cite{mostafanasab20152-dskew} develop a division algorithm in $R[x, y; \rho, \theta]$ leading to determination of generator polynomials for $2$-D skew $(\lambda_1, \lambda_2)$-constacyclic codes.

	\begin{thm}\textup{(See Mostafanasab \cite{mostafanasab20152-dskew})}
		Let $f_1(x, y), f_2(x, y) \in R[x, y; \rho, \theta]$ be two nonzero bivariate polynomials where $f_2(x, y)$ is monic. Provided that $\deg(f_1(x, y)) \geq \deg(f_2(x, y))$, there exists a pair of polynomials $h(x, y)(\neq 0), g(x, y) \in R[x, y; \rho, \theta]$ which satisfy $f_1(x, y) = h(x, y) \star f_2(x, y) + g(x, y)$ such that $g(x, y) = 0$ or $\deg(f_2(x, y)) \nleq \deg(g(x, y))(\leftarrow \deg(f_1(x, y)))$.
	\end{thm}

	\begin{lemma}\textup{(See Mostafanasab \cite{mostafanasab20152-dskew})}
		Let $C$ be a $2$-D skew $(\lambda_1, \lambda_2)$-constacyclic code in $R^\circ$ and
		$g(x, y)$ be a monic polynomial in $R[x, y; \rho, \theta]$. Then $g(x, y)$ is of the minimum degree (with respect to $``\leq"$) in $C$ if and only if $C = \langle g(x, y)\rangle_l$ .
	\end{lemma}

	\begin{thm}\textup{(See Mostafanasab \cite{mostafanasab20152-dskew})}
		Let $C$ be a $2$-D skew $(\lambda_1, \lambda_2)$-constacyclic code in $\mathbb{F}_q^\circ$ and
		$g(x, y)$ be a monic generator polynomial of $C$ in $\mathbb{F}_q[x, y; \rho, \theta]$. Then
		the following conditions hold:
		\begin{enumerate}[(1)]
			\item $g(x, y)$ divides $(x^l - \lambda_1) \star (y^s - \lambda_2)$.
			\item Suppose that $\deg(g(x, y)) = (l - k, s - t)$. Then
			\begin{center}
				$ \mathbb{B} = \left\{\begin{array}{cccc}
					g(x, y), & y \star g(x, y), & \cdots & ,y^{t - 1} \star g(x, y)\\
					x \star g(x, y), & xy \star g(x, y), & \cdots & ,xy^{t - 1} \star g(x, y)\\
					\vdots & \vdots & \ddots & \vdots\\
					x^{k - 1} \star g(x, y), & x^{k - 1}y \star g(x, y), & \cdots & ,x^{k - 1}y^{t - 1} \star g(x, y)
				\end{array}\right\}$
			\end{center}
		is a basis for $C$ over $\mathbb{F}_q$ and so $|C| = q^{kt}$.
		\end{enumerate}
	\end{thm}
	 
	\begin{thm}\textup{(See Mostafanasab \cite{mostafanasab20152-dskew})}
		Let $R$ be a domain and $\lambda_1, \lambda_2$ be units in $R$. Suppose
		that $C$ is a $2$-D skew $(\lambda_1, \lambda_2)$-constacyclic code in $R^\circ$ that is generated by
		\begin{center}
			$g(x, y) = \sum\limits_{j = 0}^{s - t - 1} \sum\limits_{i = 0}^{l - k - 1}g_{i, j}x^iy^j + \sum\limits_{j = 0}^{s - t}x^{l - k}y^j + \sum\limits_{i = 0}^{l - k - 1}x^iy^{s - t}$.
		\end{center}
		Then $g(x, y) \star xy \in C$ if and only if $g(x, y) \in R^{\rho \theta}[x, y; \rho, \theta]$.
	\end{thm}
	
	\paragraph*{} For defining two dual codes consider two $l \times s$ arrays $c = (c_{i, j})$ and $d = (d_{i, j})$ and define the product:
	\begin{center}
		$c \odot d :=  \sum\limits_{j = 0}^{s - 1}\sum\limits_{i = 0}^{l - 1}c_{i, j}d_{i, j}$.
	\end{center}
	
	Codewords $c$ and $d$ are said to be orthogonal provided $c \odot d = 0$. The dual code $C^\perp$ of a linear code $C$ of length $ls$ is the set of all $l \times s$ arrays orthogonal to all codewords of $C$. Results showing the relationship between a $2$-D skew $(\lambda_1, \lambda_2)$-constacyclic code and its dual follow.
	
	\begin{thm}\textup{(See Mostafanasab \cite{mostafanasab20152-dskew})}
		Let $|\langle \rho \rangle | \, | \, l, |\langle \theta \rangle | \, | \, s$ and $\lambda_1, \lambda_2$ be units in $R$ such that $\lambda_1 \in R^\rho, \lambda_2 \in R^\theta$. Suppose that $C$ is a code of lenght $l$s over $R$. Then $C$ is $2$-D skew $(\lambda_1, \lambda_2)$-constacyclic if and only if $C^\perp$ is $2$-D skew $(\lambda_1^{-1}, \lambda_2^{-1})$-constacyclic. In particular, if $\lambda_1^2 = \lambda_2^2 = 1$, then $C$ is $2$-D skew $(\lambda_1, \lambda_2)$-constacyclic if and only if $C^\perp$ is $2$-D skew $(\lambda_1, \lambda_2)$- constacyclic.
	\end{thm}
	
	\paragraph*{} The multiplicative set $S = \{x^iy^j \, | \, i, j \in \mathbb{N}\}$ help us to localize the ring $R[x, y; \rho, \theta]$ to the right. The elements of the localized ring $R[x, y; \rho, \theta]S^{-1}$ are of the form
	$\sum\limits_{j = 0}^{t} \sum\limits_{i = 0}^{k} x^{-i}y^{-j}a_{i, j}$, with the coefficients on right and the
	multiplication is given by $ax^{-1}y^{-1} = x^{-1}y^{-1}\rho\theta(a)$. Using this localization we notice how generator polynomials of $2$-D skew $(\lambda_1, \lambda_2)$-constacyclic codes are right divisors of $(x^l - \lambda_1) \star (y^s - \lambda_2)$ in $R[x, y; \rho, \theta]$.
	
	\begin{prop}\textup{(See Mostafanasab \cite{mostafanasab20152-dskew})}
		Let $\psi : R[x, y; \rho, \theta] \rightarrow R[x, y; \rho, \theta]S^{-1}$ be defined by  		\begin{center}
			$\psi\left(\sum\limits_{j = 0}^{t} \sum\limits_{i = 0}^{k} a_{i, j}x^iy^j\right) = \sum\limits_{j = 0}^{t} \sum\limits_{i = 0}^{k} x^{-i}y^{-j}a_{i, j}$.
		\end{center}
		Then $\psi$ is a ring anti-isomorphism.
	\end{prop}

	\begin{thm}\textup{(See Mostafanasab \cite{mostafanasab20152-dskew})}
		Let $a(x, y) = \sum\limits_{j = 0}^{s - 1} \sum\limits_{i = 0}^{l - 1} a_{i, j}x^iy^j$ and $b(x, y) = \sum\limits_{j = 0}^{s - 1} \sum\limits_{i = 0}^{l - 1} b_{i, j}x^iy^j$ be in $R[x, y; \rho, \theta]$. Suppose that $\lambda_1^2 = \lambda_2^2 = 1$. Then the following conditions are equivalent:
		\begin{enumerate}[$(1)$]
			\item The coefficient matrix of $a(x, y)$ is orthogonal to the coefficient matrix of $x^iy^j(x^{l - 1}y^{s - 1} \star \psi(b(x, y)))$ for all $i \in \{0, 1, \dots, l - 1\}$
			and all $j \in \{0, 1, \dots, s - 1\}$;
			\item The coefficient matrix of $a(x, y)$ is orthogonal to
			\begin{center}
				$\mathcal{A} = \begin{pmatrix}
					b_{l - 1, s - 1} & \theta(b_{l - 1, s - 2}) & \cdots & \theta^{s - 1}(b_{l - 1, 0})\\
					\rho(b_{l - 2, s - 1}) & \rho\theta(b_{l - 2, s - 2}) & \cdots & \rho\theta^{s - 1}(b_{l - 2, 0})\\
					\vdots & \vdots & \ddots & \vdots\\
					\rho^{l - 1}(b_{0, s - 1}) & \rho^{l - 1}\theta(b_{0, s - 2}) & \cdots & \rho^{l - 1}\theta^{s - 1}(b_{0, 0})
				\end{pmatrix}$.
			\end{center}
			and all of its column skew $\lambda_1$-constacyclic shifts and row skew
			$\lambda_2$-constacyclic shifts;
			\item $a(x, y) \star b(x, y) = 0$ in $R^\circ$.
		\end{enumerate}
	\end{thm}
	
	\begin{prop}\textup{(See Mostafanasab \cite{mostafanasab20152-dskew})}
		Let $\lambda_1^2 = \lambda_2^2 = 1, g(x, y)$ be a right divisor of
		$(x^l - \lambda_1) \star (y^s - \lambda_2)$ in $R[x, y; \rho, \theta]$ and $h(x, y) \coloneqq \frac{(x^l - \lambda_1) \star (y^s - \lambda_2)}{g(x, y)}$ with
		$\deg(h(x, y)) = (k, t)$. Suppose that $C$ is a $2$-D skew $(\lambda_1, \lambda_2)$-constacyclic
		code of length $ls$ over $R$ that is generated by $g(x, y)$. Then the bivariate skew polynomial $x^ky^t \star \psi(h(x, y))$ is a right divisor of $(x^l - \lambda_1) \star (y^s - \lambda_2)$
		and $x^ky^t \star \psi(h(x, y)) \in C^\perp$.
	\end{prop}
	
	\begin{prop}\textup{(See Mostafanasab \cite{mostafanasab20152-dskew})}
		Let $g(x, y)$ be a monic right divisor of $(x^l - \lambda_1) \star (y^s - \lambda_2)$ in $R[x, y; \rho, \theta]$ and $h(x, y) \coloneqq \frac{(x^l - \lambda_1) \star (y^s - \lambda_2)}{g(x, y)}$. Suppose that $C$ is a $2$-D skew $(\lambda_1, \lambda_2)$-constacyclic
		code of length $ls$ over $R$ that is generated by $g(x, y)$. Then for $f(x, y) \in R[x, y; \rho, \theta], \, f(x, y) \in C$ if and only if $f(x, y) \star h(x, y) = 0$ in $R^\circ = R[x, y; \rho, \theta]/\langle(x^l - \lambda_1) \star (y^s - \lambda_2)\rangle_l$.
	\end{prop}

	\section{Asymmetric quantum product codes}
	
	\paragraph*{} The paper \cite{la guardia2012asymmetric} by La Guardia provides the construction of a family of asymmetric quantum codes derived from the product code of two (classical)
	Reed-Solomon (RS) codes by applying
	the Calderbank-Shor-Steane (CSS) construction.
	
	\begin{thm}\textup{(See La Guardia \cite{la guardia2012asymmetric})}
		The product code of two Reed-Solomon codes $C_1 = [q - 1, q - \delta_1, \delta_1]_q$ and $C_2 = [q - 1, q - \delta_2, \delta_2]_q$ over $\mathbb{F}_q$ is the
		code with parameters $C_1 \otimes C_2 = [(q - 1)^2, (q - \delta_1)(q - \delta_2), \delta_1\delta_2]_q$. The Euclidean dual code $(C_1 \otimes C_2)^\perp = [(q - 1)^2, K^\perp, d^\perp]_q$, has parameters $K^\perp = q(\delta_1 + \delta_2 - 2) - \delta_1\delta_2 + 1$, and $d^\perp = \min\{q - \delta_1, q - \delta_2\}$, where $\delta_1$ and $\delta_2$ are the minimum distances of $C_1$ and $C_2$, respectively.
	\end{thm}

	\paragraph*{} Let $q_0$ be a prime power. Let $V_N := (C^{q_0})^ {\otimes N}$ denote the $N$th tensor power of $q_0$-dimensional Hilbert space $C^{q_0}$. An $[[N, k, D]]_{q_0}$ quantum code is a $q^k_0$-dimensional vector
	subspace of $V_N$ with minimum distance $D$. The connection between quantum codes and classical linear codes was established by Calderbank et al. \cite{calderbank1998quantum}. Since then, many classes of quantum codes have been constructed by using classical error-correcting codes.
	
	\begin{lemma}\textup{(See La Guardia \cite{la guardia2012asymmetric}) (CSS construction)}
		Let $C_1$ and $C_2$ denote two classical
		linear codes with parameters $[n, k_1, d_1]_q$ and $[n, k_2, d_2]_q$, respectively. Consider that
		$d_x = \min\{wt(C_1 \backslash C_2), wt(C_2^\perp \backslash C_1^\perp)\}$ and $d_z = \max\{wt(C_1 \backslash C_2), wt(C_2^\perp \backslash C_1^\perp)$\}. If $C_2 \subset C_1$, then there exists an AQECC with parameters $[[n, K = k_1 - k_2, d_z/d_x]]_q$.
	\end{lemma}

	Using the above lemma the author obtained the following

	\begin{thm}\textup{(See La Guardia \cite{la guardia2012asymmetric})}
		There exist asymmetric quantum product codes with parameters
		\begin{center}
			$[[(q - 1)^2, (q - d_1)(q - d_3) - (q - d_2)(q - d_4), d_z/d_x]]_q$,
		\end{center}
		where $d_1, d_2, d_3, d_4$ are integers satisfying the inequalities $2 \leq d_1 \leq d_2 <
		q - 1, 2 \leq d_3 \leq d_4 < q - 1$ and $d_z \geq \max\{d_1d_3, \min\{q - d_2, q - d_4\}\}, d_x \geq
		\min\{d_1d_3, \min\{q - d_2, q - d_4\}\}$.
	\end{thm}

	\begin{coro}\textup{(See La Guardia \cite{la guardia2012asymmetric})}
		There exist asymmetric quantum codes with parameters
		\begin{center}
			$[[(q - 1)^2, (q - d_1)^2 - (q - d_2)^2, d_z/d_x]]_q$,
		\end{center}
		where $d_z \geq \max\{(d_1)^2, q - d_2\}, \, d_x \geq \min\{(d_1)^2, q - d_2\}$, and $d_1, d_2$ are integers satisfying the inequalities $2 \leq d_1 \leq d_2 < q - 1$.
	\end{coro}

	\section{Conclusion}
	
	\paragraph*{} Our survey mostly covered results related to two-dimensional codes. But this theory can be easily generalized to multidimensional codes (i.e. codes formed by direct product of more than two codes). Already there have been some advances in this topic. Encoding and decoding problems for multidimensional cyclic codes along with their structure have been discussed extensively by Saints in \cite{saints1995algebraic}. Güneri and Özkaya studied the structure of multidimensional analogues of quasi-cyclic codes in \cite{guneri2016multidimensional}. Also, generalizations of various problems discussed in this survey have already been provided.  In \cite{guneri2008multidimensional}, Güneri and Özbudak extended ideas presented in \cite{guneri2004artin} to study the relationship between the weights of codewords to the number of affine rational points of Artin–Schreier type hypersurfaces over finite fields by obtaining a trace representation for multidimensional cyclic codes via Delsarte’s theorem. Andriamifidisoa, Lalasoa and Rabeherimanana have generalized Sepasdar’s method for finding a generator matrix of two-dimensional cyclic codes \cite{sepasdar2016some, sepasdar2017generator} to find an independent subset of a general multicyclic code in \cite{andriamifidisoa2019finding}. With the study of various types of codes and their properties in multiple dimensions still not completed, this is an active field of research in coding theory.

	\section{Acknowledgements}
	
	\paragraph*{} The author is grateful to Dr. Sanjay Kumar Singh for suggesting him to survey this topic which came up during a project. The author would also like to thank Dr. Nupur Patanker for her careful reading of the survey and for providing detailed comments which helped in improving the writing.

	\vspace{0.25cm}

	\paragraph*{} Indian Institute of Science Education and Research Bhopal
	\paragraph*{} E-mail address: \href{mailto:amajit16@iiserb.ac.in}{amajit16@iiserb.ac.in}

\end{document}